\documentclass[useAMS,usenatbib]{mn2e}
\input psfig.sty
\def\ltsim{\raise 2pt \hbox {$<$} \kern-1.1em \lower 4pt \hbox {$\sim$}}
\def\ltapprox{\raise 2pt \hbox {$<$} \kern-1.1em \lower 5pt \hbox {$\approx$}}
\def\gtsim{\raise 2pt \hbox {$>$} \kern-1.1em \lower 4pt \hbox {$\sim$}}
\def\gtapprox{\raise 2pt \hbox {$>$} \kern-1.1em \lower 5pt \hbox {$\approx$}}
\def\arcsec{$^{\prime\prime\,}$}
\def\arcmin{$^{\prime\,}$}

\def\rhalo{MRC~0116+111}

\def\com#1{$^\dagger$}
\def\deg{$^{\rm o}$}

\title[MRC~0116+111: a bubble-like radio halo of AGN feedback]{
A diffuse bubble-like radio-halo source MRC~0116+111:
imprint of AGN feedback in a low-mass cluster of galaxies
}
\author[Bagchi et al.]{Joydeep Bagchi$^{1}$
\thanks{Correspondence to E-mail: joydeep@iucaa.ernet.in}%\thanks{.........}, 
Joe Jacob$^{2}$,  Gopal-Krishna$^{3}$,  Norbert Werner$^{4}$, \newauthor 
Nitin Wadnerkar$^{5}$, Jaydeep Belapure$^{6}$, A. C. Kumbharkhane$^{5}$\\
$^{1}$Inter University Center for Astronomy and Astrophysics (IUCAA), Post bag 4, Ganeshkhind, Pune 411 007, India\\
$^{2}$Department  of Physics, Newman College, Thodupuzha 685 585, India\\
$^{3}$National Center for Radio Astrophysics (NCRA-TIFR), Post Bag 3, Ganeshkhind, Pune 411 007, India\\
$^{4}$Kavli Institute for Particle Astrophysics and Cosmology, Stanford
University, 452 Lomita Mall, Stanford, CA 94305, USA \\
$^{5}$School of Physical Sciences, Swami Ramanand Teerth Marathwada University, Nanded 431 606, India\\
$^{6}$Department of Physics, Pune University, Pune 411 007, India\\
}
\begin{document}

%\date{Accepted 1988 December 15. Received 1988 December 14; in original form 1988 October 11}

\pagerange{\pageref{firstpage}--\pageref{lastpage}} \pubyear{2008}

\maketitle

\label{firstpage}

\begin{abstract}
We present detailed observations of  \rhalo, revealing a luminous,  mini radio-halo 
of $\sim$~240 kpc diameter located at the centre of
a cluster of galaxies at redshift z = 0.131.  Our optical 
and multi-wavelength GMRT and VLA radio observations reveal a highly unusual radio source: 
showing a pair of giant ($\sim$100 kpc diameter) bubble-like 
diffuse structures, that are about three times larger than the 
analogous extended radio emission observed in M87 - the dominant central radio galaxy in 
the Virgo Cluster. However, in 
\rhalo\ we do not detect  any ongoing Active Galactic 
Nucleus (AGN) activity, such as a compact core or 
active radio jets feeding the plasma bubbles. The radio emitting relativistic
particles and magnetic fields were probably seeded in the past by a pair of radio-jets originating in the AGN of the central cD galaxy. The extremely 
steep high-frequency radio spectrum of the north-western bubble, located $\sim$100 kpc from cluster centre, indicates radiation losses, possibly because 
having detached, it is  rising buoyantly and moving away into 
the  putative hot intra-cluster medium. The other 
bubble, closer to the cluster centre, shows signs of
ongoing particle re-acceleration. We estimate that the radio jets which
inflated these two bubbles might have also fed  
enough energy into the  intra-cluster medium to create an enormous system of 
cavities and shock fronts, and to drive a massive outflow 
from the AGN, which could counter-balance 
and even quench a cooling flow. Therefore, this source presents an
excellent  opportunity to  understand the energetics and the dynamical  evolution of
radio-jet inflated  plasma bubbles in the hot cluster atmosphere.
%At the present moment  \rhalo\  looks like a
%`fossil' radio source, which is currently inactive and
%its twin extended lobes/bubbles of synchrotron radio plasma  not maintained
%by active jets from an AGN. 
%Another unusual aspect is detection of 
%significant polarization in 1.4 GHz NVSS (New VLA Sky Survey) data, 
%which is quite rare for a cluster radio halo.
\end{abstract}


\section{Introduction}
Multi-wavelength observations
of galaxy clusters have provided substantial evidences
for supermassive black holes (SMBH) residing in the nuclei of 
giant elliptical galaxies which  are found near the 
cluster centres and dense galactic environments.
The extremely large mass of these black holes ($M_{bh} \sim
10^{6} - 10^{9} M_{\odot}$) raises several questions: what processes 
regulate the growth of these SMBH?  why and how only some of them 
turn-on their radio jets? what is the source of  fuel for the central-engine? 
and what environmental impact could  a massive black hole and
its energetic outflows may have on  the host galaxy and the surrounding gaseous
intra-cluster medium (ICM)?  Some important clues to these 
questions are provided 
by  radio and X-ray observations of galaxy clusters which 
show that majority (as many as $\sim$70 - 100 percent) of  these 
central giant elliptical galaxies (cD galaxies) 
may become radio-loud - ejecting plasma and magnetic fields in the
form of powerful radio-jets - when they are located at the focus of
a `cooling-flow' \citep{burns90,bk94,mittal08}. The probability
of finding a central radio source increases in direct proportion to
decrease in cooling time of the inner ICM \citep{bk94,mittal08}.
A cooling-flow is established
when the central radiative cooling time ($t_{cool}$) of the ICM of a  galaxy
cluster is quite short ($t_{cool}$ \ltsim1 Gyr $<< H_{0}^{-1}$, 
the Hubble time). 
In the absence of a  central heating mechanism, the pressure balance is
disturbed and to restore the hydrostatic equilibrium, 
a steady convergent 
flow of  cooling gas towards the cluster centre is setup, in which the 
gas cools below the X-ray temperature 
and accretes onto the central elliptical galaxy \citep{Fabian94,Peterson06}. 

Strongest spectral signatures of this cooling gas are 
low energy X-ray lines below 1~keV from various 
ionization states of Iron (mainly Fe~L complex). 
Early {\it Einstein} observations of 
nearby radio-loud cD galaxies in cooling-flow clusters such as 
M87 (Virgo-A) and NGC~1275 (3C84, Perseus-A) did find the expected Fe~L
lines, thus supporting the cooling model 
\citep{canizares82}. Evidence for cooling gas was
also seen in other wavelengths, although at rates lower than
expected; $H_{\alpha}$ emitting filaments of ionized, warm gas
\citep{Heckman81}, star formation activity \citep{Mcnamara89} and
radio observations of CO molecular lines \citep{Edge01,Salome08}.

In recent years more detailed X-ray observations with {\it Chandra} and
{\it XMM-Newton} telescopes have revealed a surprising and somewhat puzzling
aspect of cooling flows; they  showed far less 
cooling below X-ray emitting temperatures than
expected, altering the previously accepted picture of 
cooling flows (cf. review by \citet{Peterson06}). Unless gas is 
thermally supported, radiative 
cooling  leads to a `cooling catastrophe', i.e., an  
inexorable inflow of cold gas onto
the central galaxy. To prevent this, some  heating mechanism was required
to raise gas temperature above $\sim 2$ keV, suppressing
the cooling flow. Although several such mechanisms were discussed, the 
most promising heating process turned out to be the energy injected into
the ICM by powerful radio jets emanating from  AGNs 
in central galaxies of clusters and groups 
(e.g., \citet{BT95,Churazov01,Churazov02,BK02}, and see
\citet{McNamara07} for a review). This AGN-heating mechanism is also found to be 
particularly effective in
suppressing the star formation activity, that one expects to be prolific
around the brightest cluster galaxies (BCGs) due to their preferential 
location in clusters. These BCGs should be accompanied by 
intense star formation activity and have blue colours, neither of 
which is observed and energy input
from AGN outbursts in clusters may be especially needed to 
explain the observations.

One of the canonical models of AGN posits accretion of gaseous 
medium as fuel for the nuclear
black hole, such that  AGN outflows are powered by
gravitational binding energy released by the
infalling matter \citep{Begelman84}. The observed strong association
of an AGN in the central galaxy and  the surrounding cooling-core  
lends good support to
this model, which suggests that the black hole `central engine' is  possibly fueled 
by accretion of cooling gas
\citep{bk94,allen06,mittal08}, with the flow rate
self-regulated by  AGN-heating \citep{Churazov01,Cattanoe07}, indicating
a complex feedback loop with tight coupling
between a central black hole and the surrounding gas of cooling core. 
Although this model is plausible and  widely used, many details of how this feedback process works remain far from clear. 

When the radio jets emerging from the central  
black hole (AGN) interact with the dense thermal plasma of the  
ICM, two bubble-like lobes of non-thermal 
plasma are inflated, which are filled with relativistic particles
and magnetic field and thus become visible in radio observations.
Such bubbles were first  proposed by \citet{GN73} and 
later identified in several clusters using radio and 
X-ray observations  (see \citet{McNamara07}).
The most clear example of this phenomenon are
the non-thermal bubbles in clusters MS0735.6+7421, Hydra-A, Abell 2052, Perseus and others, showing an
unusually large and energetic pair of radio emitting, X-ray dark cavities 
(e.g., \citet{McNamara_et_al2000,McNamara05,Blanton03,McNamara07,Wise07}).
Episodic (on-off) activity of radio jets injects non-thermal radio bubbles which 
may heat the ICM via weak shocks, and additionally these  plasma bubbles are  responsible for the mechanical (PdV) work done on the ICM for heating it,
which is one of the favored mechanism of AGN-ICM feedback. 
%In clusters these AGN driven, radio emitting non-thermal 
%plasma bubbles are responsible for the mechanical $pdV$ work done on the ICM, 
%which is the favored mechanism for heating of ICM by AGN activity.
%Spectacular examples of such
%feedback process in action are the radio lobe inflated

Several examples of these bubble and cavities, with diameter 
ranging from 1-100 kpc, have been found in the hot 
atmosphere surrounding galaxies in
groups and clusters (e.g., \citet{Birzan04,Dunn_fabian04}). In
few sources the energy involved is $\sim 10^{60-62}$ erg, the most
powerful radio outbursts known (e.g., \citet{McNamara05,McNamara07}). The
energy involved is large enough to strongly affect or even quench any
cooling flow, and to drive large-scale outflows that redistribute and heat
the gas on cluster-wide scales. The radio lobe plasma fill the cavities, 
which shows that expanding jets have displaced
the ICM thermal gas, excavating X-ray dark cavities and possibly shock heating the
surrounding gaseous medium (e.g., \citet{Fabian03,McNamara_et_al2000,
McNamara05}). This heat input into the ICM may be large enough to counter-balance 
the cooling loss, and may even stop the accretion flow of matter onto the SMBH, thus starving the central engine of its fuel. Therefore,  bubble-like diffuse  
radio sources residing near  cluster centres  are extremely  
effective  probes of the poorly understood 
physics of the radio galaxy-ICM feedback process. 

Also directly related to these issues is the  detection of  diffuse non-thermal
synchrotron radio plasma and magnetic fields mixed with the ICM of clusters. 
Approximately 30 galaxy clusters show {\it giant radio haloes} -- 
centrally located synchrotron radiation extended on mega-parsec 
scales with a regular morphology, similar to that observed in X-rays. 
A few galaxy clusters also show
peripherally located {\it radio relics} which have elongated morphology and that
may extend over 100 kpc -- 1 Mpc scales. The radio emission from giant radio haloes
and radio relics is not directly connected to the 
galaxies in clusters and both are  mainly observed in 
merging clusters, which indicates
that radio emitting relativistic electrons may 
have been accelerated in-situ in merger 
induced shocks and/or  turbulence (cf. review \citet{Ferrari08} and references therein). In the case 
of relics, an alternative model has been proposed, which invokes adiabatic compression of relic radio lobes by cluster merger shocks \citep{EGK01}. 

A third remarkable category of large scale 
diffuse radio sources are the {\it radio 
mini-haloes}. Unlike the mega-parsec scale  giant radio haloes
and relics (which are found to be associated with cluster merger
phenomenon and hence rarely found in cooling core clusters), radio
mini-haloes are not only a few times smaller in size ($\sim$100 - 500 kpc, 
comparable to size of the cooling core) 
but they are seen to surround the central radio galaxies of some cooling core
clusters (see reviews: \citet{FG08}; \citet{Ferrari08}).
Radio mini-haloes are rare objects,  have low surface brightness and 
a steep spectral index ($\alpha$\ltsim\,-1)\footnote
{We use definition: flux density$(S_{\nu})$ $\propto$ frequency$(\nu)^{\alpha}$}.
The fact that mini-haloes are mostly observed at the centres of cooling-core clusters indicates that their
origin and evolution are  connected to the energy feedback from the central AGN 
into the ICM via radio jets and cooling/heating processes. 
Thus, while mini-haloes share with gaint-haloes and relics the properties
of a steep radio spectrum and very low surface brightness, their physical
origins are probably different. This is corroborated by the finding
that unlike the mini-haloes, giant haloes have remarkably similar synchrotron
emissivities, albeit the two are similar morphologically \citep{Murgia09}.
Radio mini-haloes are still a very poorly understood phenomenon and in 
observational terms, they being roughly centred at the
dominant cluster radio galaxy makes their detection particularly
challenging. Three early examples of radio mini-haloes are associated
with the cooling flow clusters: Perseus-A \citep{pedlar90}, 
A~2142 \citep{Giovannini_Feretti2000} and RX~J1347.5-1145
\citep{Gitti_et al2007}. Continued searches have so far revealed no more
than 10 mini-haloes (see \citet{Ferrari08}; \citet{FG08} and references therein).
Due to  their very steep spectrum, radio-haloes in clusters
are difficult to detect at higher frequencies,  and therefore a low
frequency radio telescope  such as the GMRT is well suited for their 
study (e.g., \citet{Brunetti08};\citet{Venturi08}).

The focus of present paper is on a diffuse halo-like  radio source \rhalo\  
and its radio and optical properties, as revealed in our detailed 
observations, their astrophysical implications and impact on the above topics.
Earlier a brief report on this  radio source  
matching the characteristics
of a mini-halo, was presented by some of us, based on
VLA and GMRT observations \citep{IAU2002}. 
These radio observations revealed
an amorphous source of an ultra-steep radio spectrum $(\alpha \sim -1.3 )$
 and size $\sim$ 1.5\arcmin.  This {\it Molonglo Reference 
 Catalogue} source MRC~0116+111, had
been resolved on arc-minute scale in the {\it Ooty Lunar Occultation Survey}
at 327~MHz \citep{Joshi_Singal_1980}.
An R-band image taken with the ESO {\it New Technology Telescope} (NTT) showed
that this amorphous radio source is centred at the dominant member of a
galaxy group having at least 3 - 4 members
for which NTT/EMMI spectroscopy yielded redshifts close to 0.131
\citep{IAU2002}. More recently, this system has been listed
in a catalogue of distant  clusters of galaxies \citep{Lopes04}. 
In order to clarify the nature of this peculiar radio source and its
relation to the brightest cluster galaxies, we have imaged it freshly
with GMRT at 240 MHz, 621 MHz and 1.28 GHz with high sensitivity and resolution, 
and have also carried out an improved
analysis of the existing VLA snapshot data at 1.4 and 4.8 GHz. Additionally, we
have taken its deep {\it B,V,R,I\,} CCD images with the 2-mt optical  
telescope located at the IUCAA Girawali Observatory (IGO), Pune, India. 

%Due to  very steep spectrum of radio halos in clusters, they 
%are difficult to detect at high frequencies  and a dedicated low 
%frequency radio telescope such as GMRT is ideally suited for their study.
%Recently \citet{Brunetti08} discovered a very steep spectrum 
%giant radio halo in Abell 521 galaxy
%cluster from 240 MHz GMRT observation, which supports its origin in turbulent
%Fermi-type acceleration of primary electrons. 

In this work we have used a Hubble constant $H_{0}=$ 70~km/(s\-\ Mpc), 
and $\Lambda$-CDM `concordance' cosmology with
$\Omega_{M}=0.27$ and $\Omega_{\Lambda}=0.73$, which results in 
a luminosity distance $D_{L} = 617.2$ Mpc, and linear scale of 
2.34 kpc per arcsec for a redshift $z = 0.131$ of \rhalo.
%%%%%%%%%%%%%%%%%%%%%%%%%%%%%%%%%%%
%%%%%%%%%%%%%%%%%%%%%%%%%%%%%%%%%%%%%%%%%%%%%%%%%%%%
\begin{figure}
\psfig{file=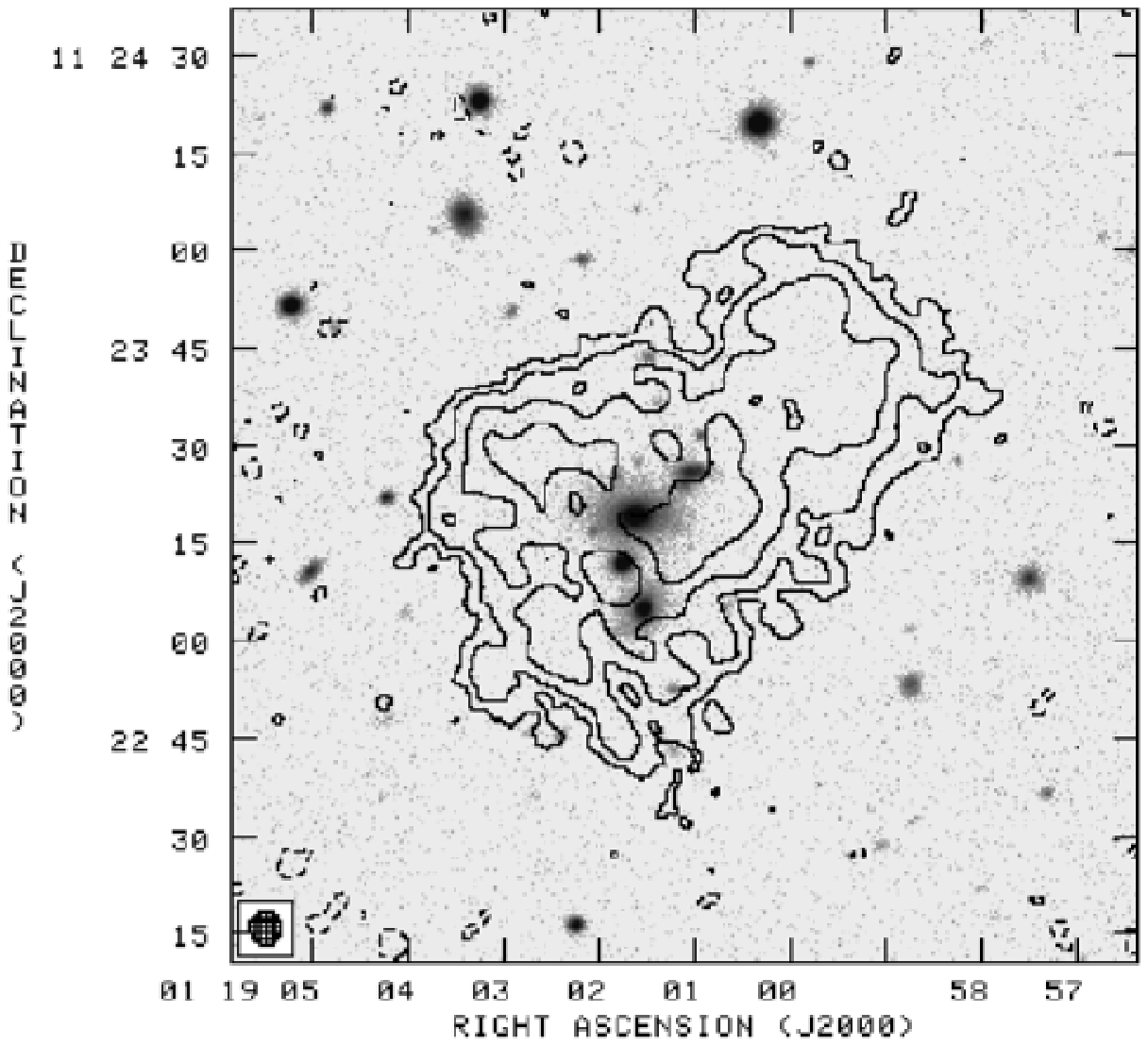,width=8.3truecm}
\psfig{file=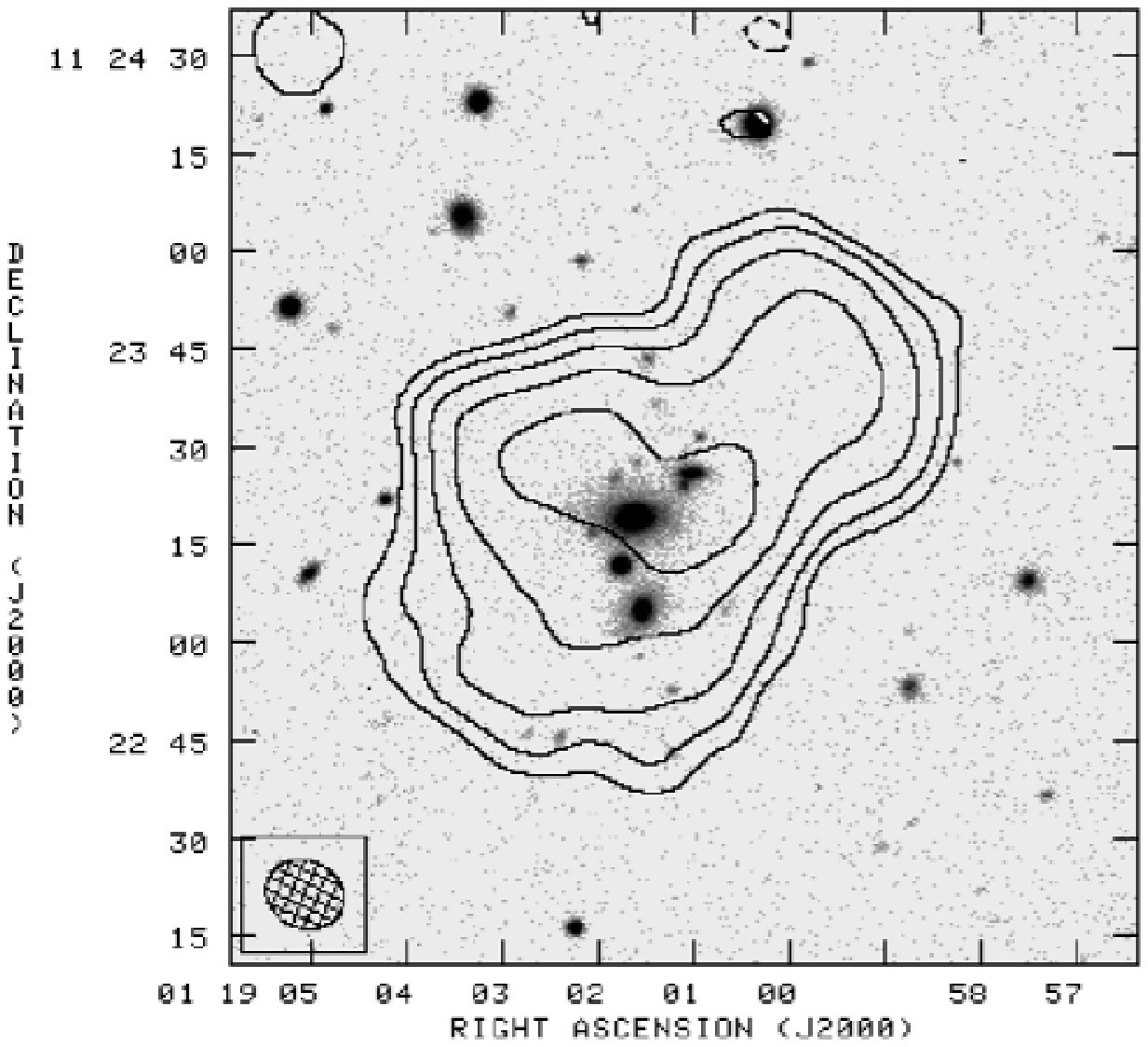,width=8.3truecm}
\caption{{\bf Upper panel:} GMRT 1.28 GHz map of \rhalo\ shown with 
contours (levels: $\pm$0.24, 0.48, 0.96, 2, 4, 8 mJy/beam, noise 
r.m.s. 0.08 mJy/beam, beam: 5\arcsec FWHM circular, plotted inside box)
overlayed on IGO R-band image. No AGN (radio core) is visible down to $\sim$~1 mJy/beam flux density limit, and no radio jets or lobes are detectable either.
{\bf Lower panel:} GMRT 240 MHz map shown with
contours (levels: $\pm$4, 8, 16, 32, 64, 128 mJy/beam, noise
r.m.s. 1.35 mJy/beam, beam: 12.15\arcsec$\times$10.31\arcsec\, FWHM at
65.4\deg\, PA,  plotted inside box). In the background is shown the
IGO R-band image of the galaxy cluster.}
%(lower image) GMRT 621 MHz false-colour image  (noise r.m.s. 0.15 mJy/beam, beam: 6\arcsec circular) showing the twin bubble-like diffuse 
%radio structure. The dotted lines delineate the two radio bubbles with 
%projected dimensions $\sim$27.5\arcsec $\times$17.5\arcsec 
%(semi-major and minor axes 
%of ellipse) for the smaller elliptical north-west 
%bubble and $\sim$35\arcsec (circular radius) 
%for the spherical larger bubble south-east of the cluster center.}
\label{gmrt_1}
\end{figure}

\section[]{Observations and data analysis}
\subsection{Radio observations}

Multi-frequency  {\it Giant Meterwave Radio
Telescope} (GMRT)\footnote {The GMRT is a national facility operated
by the National Centre for Radio Astrophysics of the TIFR, India}
observations were carried out on December 1 and 2, 2007 
at 1.28 GHz ($\lambda$ = 23.4~cm), 
621 MHz ($\lambda$ = 48.3~cm) and 240 MHz ($\lambda$ = 125~cm) 
using the  128 channel FX correlator. GMRT is an earth-rotation aperture-synthesis
array of 30 telescopes operating in 150 - 1420 MHz range  \citep{Swarup1991}.
The data were recorded in the spectral-line mode with 128 channels per
sideband, resulting in a spectral resolution of 125 kHz per channel.
The 621 and 240 MHz observations were
taken simultaneously with the co-axial dual-band feed system of GMRT.
At 240 MHz we used a bandwidth of only 6 MHz
out of total 16 MHz available in one sideband (USB), and at 621 MHz full bandwidth of 16 MHz in one sideband (USB) was used.
At 621 MHz only the right-handed circular polarization (RR)  and
at  240 MHz  only the left-handed circular polarization (LL) was recorded.
Both right and left circular polarization (RR, LL) data were recorded at 1.28 GHz, 
and bandwidth of 16 MHz was used in one sideband (USB).
%The data were recorded in the spectral-line mode with 128 channels per
%sideband, resulting in a spectral resolution of 125 kHz per channel.
During each observing run a
sequence of $\sim$20 to 30-minute scans of the source were 
taken, interspersed
with the VLA phase calibrator 0204+152 (4C~+15.05). 
The telescope gain, bandpass response
and flux density scale 
were calibrated  by observing  the
primary flux density calibrator  3C~48,  and boot-strapping the flux density 
scale to the standard `Baars-scale' \citep{Baars77}.
After excising  faulty data and RFI, mainly affecting
low frequency observations, the calibrated data were
transformed into the image plane using the standard routines in  AIPS,
including a few cycles of self-calibration. By applying suitable tapers
to the visibilities,  final maps with synthesized Gaussian beams of FWHM
ranging 5 to 6\arcsec\, at 1280 and 621 MHz, 
and $\sim$10\arcsec\, at 240 MHz were obtained (see Fig.~\ref{gmrt_1} and
Fig.~\ref{gmrt_2}).

The first of {\it Very Large Array}\footnote {The 
Very Large Array (VLA) is a facility of the National Radio Astronomy
Observatory (NRAO). The NRAO is a facility of the National Science
Foundation, operated under cooperative agreement by 
Associated Universities, Inc.} observations 
took place on June 22, 1992 in 
C-band (DnC array configuration, 4.88 GHz and 4.83 GHz centre frequencies, 
using two IFs of 50 MHz BW each). The total integration time on
\rhalo\  was about 10 minutes and VLA calibrator B0235+164 was
used for phase calibration. VLA calibrator 3C~48 was also observed for
amplitude calibration. The second VLA observation was made 
on June 12, 1997 in L-band (CnB  configuration, 
1.46 GHz and 1.38 GHz centre frequencies, using two IFs of 50 MHz BW each).
The total on-source integration time  was about 38 minutes and 
VLA flux calibrator 3C~48 was used for both amplitude
and phase calibrations. For data reduction we used standard routines 
available in AIPS software. Further editing, and imaging with a 
few cycles of self-calibration resulted in good
quality radio maps. After deconvolution, the final
maps were restored with synthesized Gaussian beams of FWHM
12\arcsec\  at both the frequencies (see Fig.~\ref{gmrt_3}).
Table~1 gives the details of observing runs with the
different radio telescopes and the sensitivities and resolutions 
that were achieved at various observing frequencies.

For obtaining a spatially resolved spectral index information, 
we combined  GMRT 240 and 621 MHz images
for a low frequency spectral index map, and VLA 1.4 and 4.8 GHz
images for a high frequency spectral index map. These are shown in 
Fig.~\ref{maps_spectindx}. For this purpose, radio data of 
\rhalo\  were specially processed (if required, by giving
suitable tapers in the visibility plane) so  
as to obtain a  uniform  
angular resolution of 12\arcsec\ (FWHM beam size) at all the
frequencies. When combining
the radio images we have included  only the pixels with
amplitudes $\sim3.5$ times 
above the  r.m.s. noise, to prevent spurious structures 
from appearing in the spectral index maps. These  cut-offs  were
given at values 4 mJy/beam, 0.5 mJy/beam, 0.27 mJy/beam and 0.15 mJy/beam
on the 240, 621, 1425 and 4860 MHz maps respectively. 

\begin{table*}
%\centering
\begin{minipage}{175mm}
\caption{Details of GMRT and VLA observations of \rhalo}
\begin{tabular}{@{}lrrrrrrrrrr@{}}
%\begin{tabular}{lrrrrrrrrrr}
\hline
\hline\noalign{\smallskip}
Frequency (MHz)    &&240 && 621 &&1280 && 1425 &&4860 \\
\hline
Telescope      &&GMRT&&GMRT&&GMRT&&VLA&&VLA \\
% Phase center\\
% RA(2000)              &&01$^h$19$^m$01$^s$&&Same&&Same&&Same&&- \\
% DEC(2000)             &&11$^\circ$23$^m$23$^s$&&Same&&Same&&Same&&- \\
Date of Observation    &&1 Dec. 2007&&1 Dec. 2007&&2 Dec. 2007&&12 June 1997&&22 June 1992  \\
%Phase calibrator &&xx && xx &&xx && xx &&xx \\
Flux Calibrator &&3C 48 && 3C 48 &&3C 48 && 3C 48 &&3C 48 \\
Bandwidth (MHz)        &&6&&16&&16&&50&&50 \\
Integration time per visibility &&16.9 s&&16.9 s&&16.9 s&&10 s&&10 s\\
FWHM synthesized beam &&12.15\arcsec$\times$10.31\arcsec&& 6\arcsec circular &&5\arcsec circular && 12\arcsec circular &&12\arcsec circular \\
RMS map noise (mJy/beam) &&1.35&&0.15&&0.08&&0.08&&0.04 \\
Integrated flux density (Jy) &&$1.15\pm0.05$&&$0.47\pm0.01$&&$0.19\pm0.01$&&$0.140\pm0.005$&&$0.035\pm0.005$ \\
\hline
%Calculated Properties: && &&  && &&  && \\
% \\
%Spectroscopic redshift (central cD galaxy)      && 0.131 && && && && \\
%621 MHz Radio Luminosity ($L_{\rm 621\,MHz}$)       &&$1.21 \times 10^{25}$ Watt/Hz && && && && \\
%1400 MHz Radio Luminosity ($L_{\rm 1.4\,GHz}$)        &&$4.57 \times 10^{24}$ Watt/Hz && && && && \\
%Bolometric Radio Luminosity ($L_{\rm radio}$)   &&$3.64 \times 10^{34}$ Watt&& && && && \\
%(over 10 MHz - 10 GHz range) && &&  && &&  && \\
%`break' frequency in spectrum &&$\sim400$ MHz &&  && &&  && \\
\hline
\end{tabular}
\end{minipage}
\end{table*}

\begin{table*}
%\centering
\begin{minipage}{175mm}
\caption{Observed and calculated parameters}
\begin{tabular}{@{}lrrrrrrrrrr@{}}
%\begin{tabular}{lrrrrrrrrrr}
\hline
\hline\noalign{\smallskip}
%Calculated Properties: && &&  && &&  && \\
	% \\
	Spectroscopic redshift (central cD galaxy)      && 0.1316 && && && && \\
	Apparent magnitudes$^{\dagger}$ (central cD galaxy)      &&$m_{B} = 18.79$, $m_{V} = 17.69$, $m_{R} = 16.69$ && && && &&\\
	621 MHz Radio Luminosity ($L_{\rm 621\,MHz}$)       &&$1.21 \times 10^{25}$ W Hz$^{-1}$ && && && && \\
	1400 MHz Radio Luminosity ($L_{\rm 1.4\,GHz}$)        &&$4.57 \times 10^{24}$ W Hz$^{-1}$ && && && && \\
	Bolometric Radio Luminosity ($L_{\rm radio}$)   &&$3.64 \times 10^{34}$ W&& && && && \\
	(over 10 MHz - 10 GHz range) && &&  && &&  && \\
	`Break' frequency in spectrum &&$\sim400$ MHz &&  && &&  && \\
	Elliptical North-west bubble dimension &&$\sim$27.5\arcsec $\times$17.5\arcsec (= 64.4 kpc
		$\times$41 kpc)&&  && &&  && \\
	    (projected semi-major and minor axes)&& &&  && &&  && \\
	    Circular South-east bubble dimension &&$\sim$35\arcsec (= 82 kpc) &&  && &&  && \\
	    (projected circular radius)&& &&  && &&  && \\
	    \hline
	    \hline
	    \end{tabular}\\
$\dagger$ Corrected for galactic extinction (NASA Extragalactic Database).
	    \end{minipage}
	    \end{table*}

\subsection{Optical imaging and spectroscopic observations}
The diffuse radio-halo source \rhalo\  was earlier identified by us to be
centred on the  dominant (cD-like) galaxy of a poor group with 3-4 members 
\citep{IAU2002}. Recently, at the position of this radio source a relatively poor 
galaxy cluster NSCS~J011904+112133 was identified
by \citet{Lopes04} in their  survey of clusters from DPOSS-II photographic
plates, with  a  photometric redshift estimate  0.16. However, From our 
NTT spectroscopy  we  placed the cluster at 
a redshift of 0.131 (see \citet{IAU2002} and below). 
Other than this no other optical or 
spectroscopic data was found in the literature. 
We carried out optical broadband ({\it B,V,R,I})  imaging  observations of the
host cluster using the recently commissioned 
2-mt telescope at the  IUCAA  Girawali 
Observatory (IGO), located about 80~km from Pune, India. 
The observations were taken on the night of 31st December 2007
using the {\it IUCAA Faint Object Spectrograph and Camera} (IFOSC) mounted at
the Cassegrain focus.
IFOSC employs an  EEV 2Kx2K, thinned, 
back-illuminated CCD with 13.5 $\mu$m 
pixels. The spatial sampling scale at the detector is 44$\mu$m arcsec$^{-1}$ 
giving a field of view of about 10.5 arcminutes on the side. 
During observation the seeing conditions were good, but with 
fair to occasionally poor sky transparency.

\begin{figure}
\psfig{file=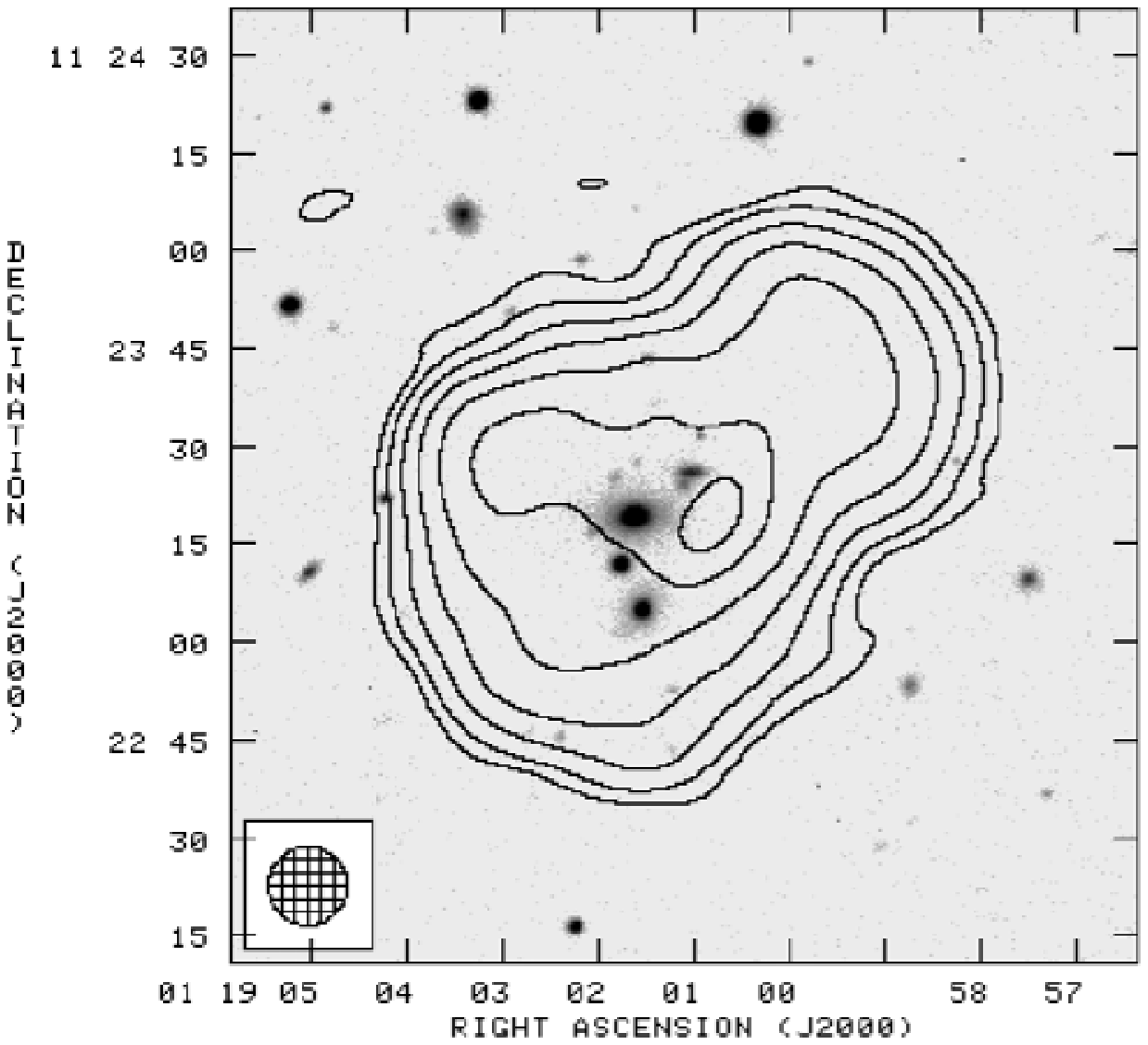,width=8.3truecm}
\psfig{file=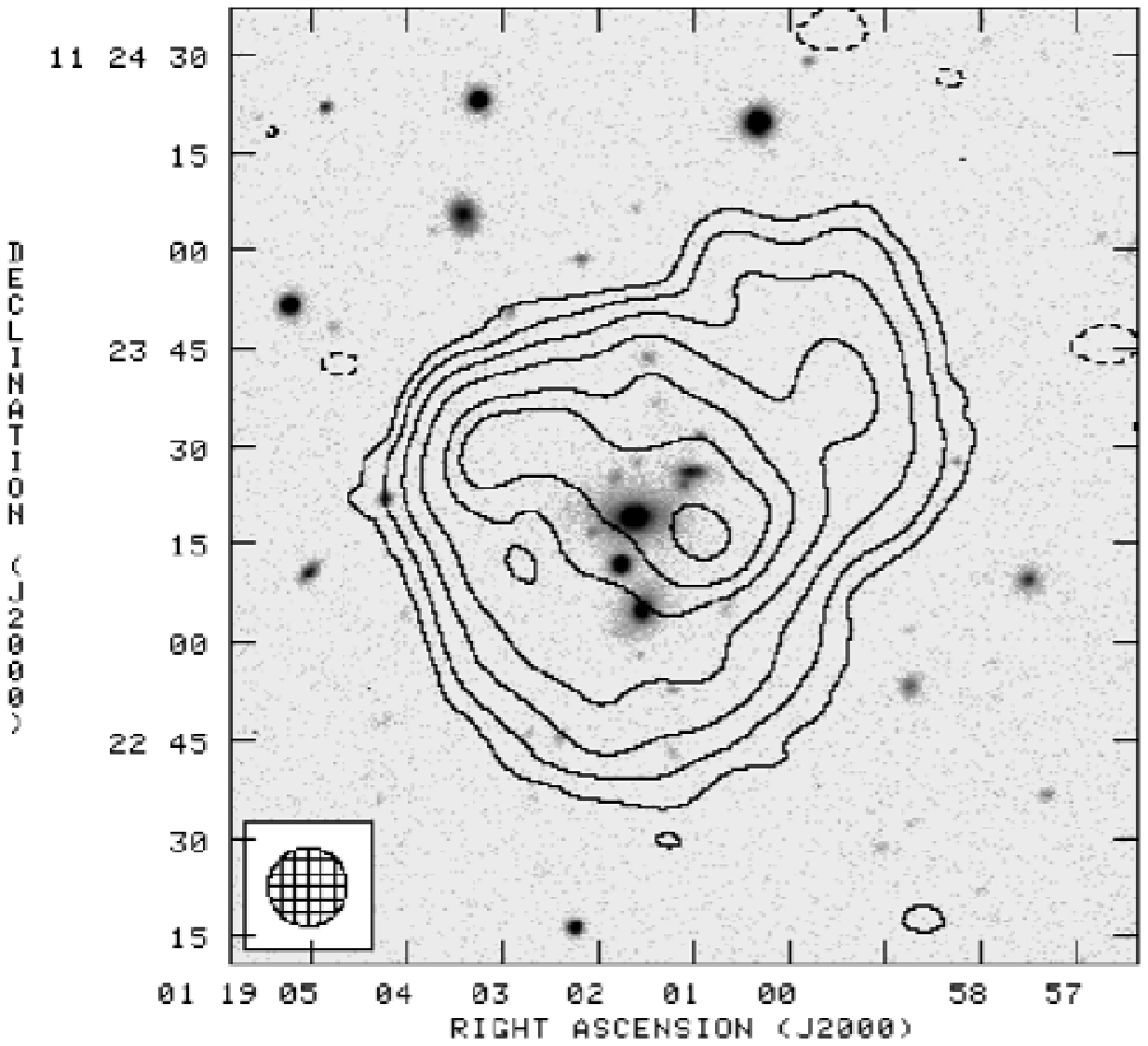,width=8.3truecm}
\caption{VLA radio maps of \rhalo\ overlayed on
IGO R-band optical image of the galaxy
cluster. The upper panel shows L-band (1425 MHz) map
drawn with contour levels $\pm$0.24,0.48,1,2,4,8 and 12 mJy/beam and 
lower panel shows the C-band (4860 MHz) map drawn with contour 
levels $\pm$0.2,0.4,0.8,1.6,2.2 and 3 mJy/beam. The resolution 
is 12\arcsec\ for both the images, shown by the FWHM of the beam  
plotted in the lower left corner.}
\label{gmrt_3}
\end{figure}

\begin{figure}
\centerline{\psfig{file=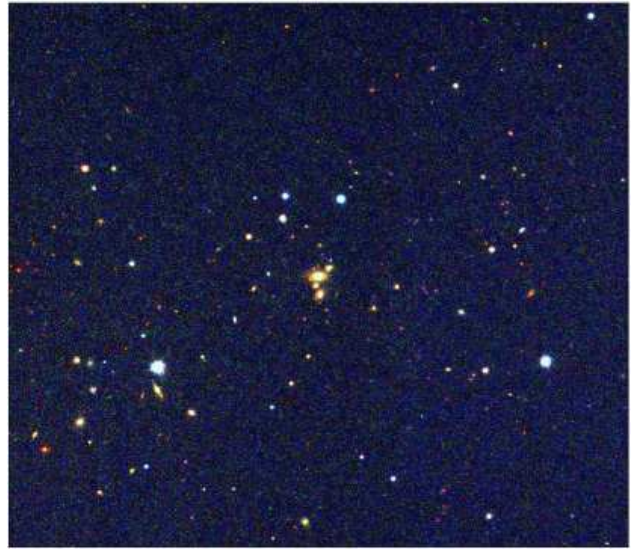,width=8.3truecm}}
\caption{ A {\it B,V,R}  false-colour
composite of the central region of galaxy
cluster hosting \rhalo.  The region shown is about 8.3\arcmin$\times$7\arcmin in size 
and  north is on top and east is to the left. This image is made using the IUCAA  Girawali
Observatory (IGO) 2-meter telescope.}
\label{IGO_BVR}
\end{figure}

We took several frames in each filter with total exposure time
sufficient to achieve adequate photometric accuracy. Photometric zero points
and colour transformations were defined by observing Landolt's
standard field {\it Rubin~149}  \citep{Landolt}. The analysis of the CCD frames
was done in a standard way using the IRAF software. The
{\it I}-band frames suffered from strong fringing, which could 
not be completely removed by flat fielding; therefore we report 
only the   {\it B,V} and {\it R} band
data in this paper. A composite {\it B,V,R}  colour image of the host 
galaxy cluster is shown in Fig.~\ref{IGO_BVR}. Several elliptical 
galaxies are visible within the contours of GMRT and VLA radio maps 
(Fig.~\ref{gmrt_1} and \ref{gmrt_3}) 
with a giant cD-like galaxy of $m_{v} = 17.69$ mag ($M_{v} = -21.38$ mag)
seen in the centre at the
position RA = 01$^h$19$^m$1.69$^s$,
Dec.= +11$^{\rm o}$ 23$^{\rm '}$18.4$^{\rm ''}$ (J2000).

\begin{figure}
\centerline{\psfig{file=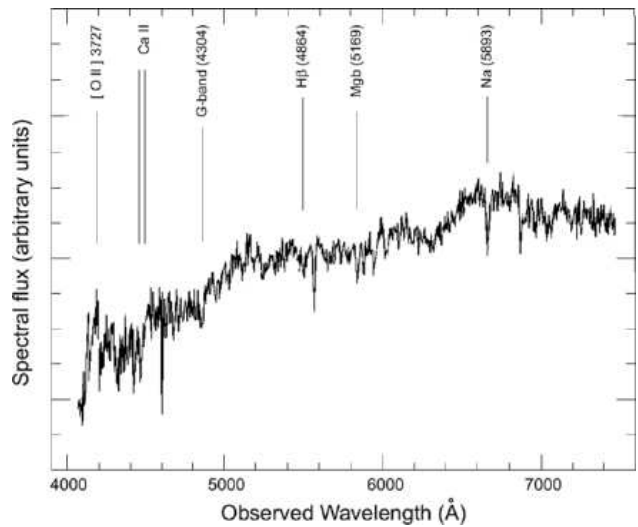,width=8.3truecm}}
\caption{ EMMI ({\it ESO Multi-Mode Instrument}) low resolution (spectral resolution 3.7\AA) slit-spectrum of
the central cD galaxy in \rhalo\ cluster taken with grism-3 optics on the New Technology Telescope (NTT).
The abscissa shows observed wavelengths and  ordinate, the spectral flux on an
arbitrary scale. The prominent spectral lines are marked along with their rest
wavelengths. The spectrum gives a redshift $z=0.1316$ for this galaxy.
}
\label{NTT_spect}
\end{figure}

For spectroscopy we used  the ESO/La~Silla  3.6-meter
New Technology Telescope (NTT) and EMMI ({\it ESO Multi-Mode Instrument}). 
A 2$\times$10 minutes low resolution slit-spectrum (slit width 1.5\arcsec, spectral resolution 3.7\AA) was taken with the 
grism-3 optics on the EMMI (October 1991) which
gave a redshift $z=0.1316$ for the 
brightest central cD galaxy (Fig.~\ref{NTT_spect}), based on the absorption 
lines of Na ($\lambda$5893), 
Mgb ($\lambda$5169), H$\beta$ ($\lambda$4861), G-band ($\lambda$4304), and the Ca~H,K break. 
A probable weak [0II] ($\lambda$3727) emission line is visible  at the left-most edge of the spectrum, 
where the S/N ratio is comparatively lower. The second brightest 
elliptical galaxy $\sim15$\arcsec\ south of cD (Fig.~\ref{IGO_BVR}) was found to have a close redshift $z=0.1309$,
suggesting a group-like environment around the central cD.

\begin{figure}
\centerline{\psfig{file=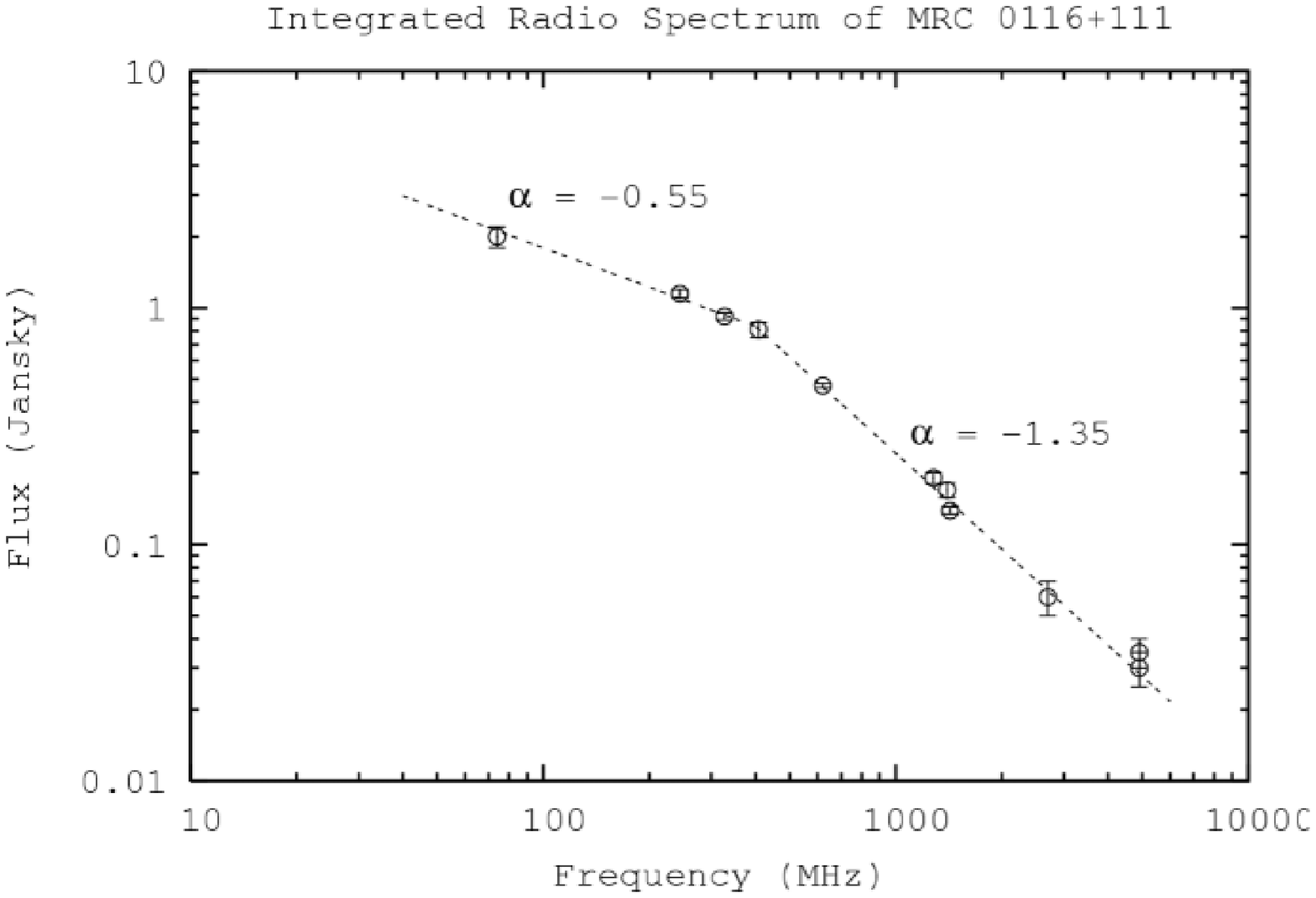,angle=-90,width=8.3truecm}}
\caption{The integrated radio spectrum of \rhalo\  between
74 MHz and 4.9 GHz. Piecewise linear
least-square  fit to the data points shows a
sudden spectral steepening beyond
the `break' frequency $\nu_{br} \sim$400 MHz. The low-frequency spectral
index  is $\alpha_{l}=-0.55(\pm0.05)$ and  the
high-frequency spectral index is $\alpha_{h}=-1.35(\pm0.06)$. The
GMRT and VLA observed
flux density values are listed in Table~1. Other data points are
taken from literature as follows:
74~MHz [VLSS: \citet{vlss}], 327~MHz [\citet{IAU2002}],
408~MHz [Molonglo: \citet{molonglo}], 1400~MHz [NVSS: \citet{nvss}],
2700~MHz [Effelsberg: \citet{IAU2002}] and
4.8~GHz [GreenBank: \citet{87gb}].
}
\label{rad_spectr}
\end{figure}

\section{Physical picture of the radio source}

\subsection{A twin bubble-like diffuse radio-halo morphology: A `quiescent' 
radio galaxy?}
\rhalo\  shows a highly diffuse `halo' or `bubble' like structure which  
bears close resemblance to the amorphous structure of so-called 
`radio mini-haloes' (see Sect.~1 and \citet{Ferrari08}). 
%Mini-halos are rarely seen,  low surface brightness amorphous radio sources 
%with a steep spectral index ($\alpha$\ltsim\,-1),
%which are found around powerful radio galaxies at the center of cooling-core
%clusters \citep{Ferrari08}. In addition to `radio relics' and giant-halos, 
%RMHs represent another new class of radio emission associated with ICM.
%Their physical size is usually in
%the range $\sim$100 - 500 kpc, comparable to size of the cooling core
%of ICM. In contrast, the giant radio-halos are Mpc-scale 
%diffuse radio sources (not directly related to AGNs)
%found in the central regions of galaxy clusters which usually show 
%signs of recent mergers and no central cooling flows \citep{Ferrari08}.
The physical size, radio spectral index and morphology of \rhalo\
are broadly consistent with the properties of a cluster radio mini-halo.
In Table.~3  we compare  the physical properties of \rhalo\  with a few other well
known radio mini-haloes, including the well known radio galaxy Virgo-A (M87). We note that the radio luminosity of \rhalo\  at 
1.4 GHz, $L_{\rm 1.4\,GHz}$ = $4.57 \times 10^{24}$ W~Hz$^{-1}$, is quite
large, comparable to the luminous radio-haloes in Perseus and A2390 clusters. 
%Its radio luminosity at 621 MHz, $L_{\rm 621\,MHz}$ = $1.21 \times 10^{25}$ %W~Hz$^{-1}$, 
Also, its bolometric radio luminosity (over 10 MHz - 10 GHz range), 
$L_{\rm radio}$ = $3.64 \times 10^{34}$ W, close to the divide between 
FRI and FRII class radio sources, would place it amongst the most luminous radio haloes known.

GMRT 621 MHz radio map of \rhalo\  (Fig.~\ref{gmrt_2}) 
and its other radio maps (Fig.\ref{gmrt_1}, \ref{gmrt_3}) clearly show 
that on larger scales ($\sim$100 kpc) the radio-halo  emission  
comprises  of two  diffuse  
`lobe'-like structures  surrounding the  central group of galaxies, 
which we suggest are actually a pair of giant, radio emitting  
plasma bubbles, filled with relativistic
particles (cosmic rays) and magnetic field. In Figure~\ref{gmrt_2} the 
dotted lines delineate this  pair of 
radio bubbles,  having the projected dimensions 
$\sim$27.5\arcsec $\times$17.5\arcsec (= 64.4 kpc 
$\times$41 kpc) - the semi-major and minor axes
of ellipse for the smaller elliptical north-west
bubble,  and $\sim$35\arcsec (= 82 kpc) radius
for the larger spherical bubble to south-east of the cluster centre.  

The total spatial extent of  
radio emission is $\sim$240 kpc which apparently surrounds the  dominant
 brightest cD-like elliptical galaxy of the cluster (Fig.~\ref{IGO_BVR}). 
%at RA = 01$^h$19$^m$1.696$^s$, 
%Dec.= +11$^{\rm o}$ 23$^{\rm '}$18.36$^{\rm ''}$ (J2000).
However, based on  the detection sensitivities achieved in our present 
GMRT and VLA observations (see Table~1),
there is no strong evidence that the extended radio emission  
emanates from  this or any  
other galaxy enclosed within the radio contours. The 
highest resolution GMRT 1.28 GHz map (Fig.~\ref{gmrt_1}) and   
other radio maps have not detected a compact active radio nucleus (AGN) and 
radio-jets directly feeding into the 100-kpc scale twin plasma lobes/bubbles of \rhalo.  In this respect \rhalo\ is very unusual and it differs from  other well known mini haloes, including canonical M87 and Perseus-A \citep{Ferrari08,Cassano08,OEK,pedlar90}, which  have 
detectable powerful AGN cores and jets. The high resolution GMRT 1.28 GHz image of \rhalo\  also do not show any other compact AGN-related background sources blended  with the radio-halo emission, which clearly rules out its origin from such  a contamination (Fig.~\ref{gmrt_1}).

Therefore, this suggests to us a physical 
picture in which the relativistic
particles and magnetic fields of \rhalo\  halo were initially seeded by a 
pair of radio-jets originating in AGN activity in the past, most likely from the nuclear SMBH of the central cD galaxy, as revealed by the
pair of plasma bubbles straddling this galaxy. However,  at present
\rhalo\ resembles a 
quiescent or `fossil' source -- in the sense that
its twin extended lobes/bubbles of  radio plasma  not energized
by active jets from an AGN -- as observed in classical radio galaxies. 
This suggests that the SMBH embedded in the nucleus of the cD galaxy, 
which was once active, probably has turned-off its radio jets sometimes 
back in the past. Only a handful of such `fossil' or `dying' radio galaxies
are known so far, which are more easily detected at 
low radio frequencies due to their very steep spectra (e.g., \citet{Parma07,Jamrozy04}). It is possible that we
have caught the central massive black hole, which created  \rhalo, 
during a low or even quiescent state of its  activity cycle.  
If the heat input into the ICM by these expanding radio bubbles is 
large enough to significantly offset the cooling
loss, the flow of accreting matter onto the SMBH may stop, 
thus starving the `central engine' of its fuel. So we expect 
that this  radio bubble-fed energy exchange 
should give rise to an episodic triggering of the AGN itself, 
establishing a self-regulated activity pattern between cooling flow and AGN activity. 
%The duty cycle of radio galaxies is estimated to be -------.
We point out that this cyclic process, even if in operation, 
does not imply that some form of  {\it in situ} re-acceleration 
of particles in shocks or turbulence  does not take place in the 
central cooling core and the extended lobes of \rhalo\  -- a scenario, in fact 
supported by both the  integrated radio spectrum  and the spectral 
index maps discussed below.

\subsection{Possible location of  \rhalo\ in a low-mass poor-cluster/group environment}

 According to our present understanding of the origin and
evolution of mini-haloes, their diffuse, steep-spectrum halo or bubble
like morpholgy results from strong interaction of  radio jets with the
surrounding dense and cool gaseous medium of the cluster core.
At present  there are no available good X-ray data for this system
and therefore we lack information about the physical state of ICM  in the
core of cluster hosting \rhalo. Such data are required for
quantifying accurately the energy input rate by the AGN jets into the
plasma bubbles of mini-halo as well as their dynamics and interaction
with the ICM. The X-ray luminosity  is also a good indicator
of the optical richness and mass of the host system.  We find that
the mini radio-halo \rhalo\  is located in an under-exposed region of {\it ROSAT
All Sky Survey} (RASS), which only allows us to determine an upper limit of
$<$1.4$\times$ 10$^{43}$ erg s$^{-1}$ for its X-ray luminosity in 0.5 -- 2 keV band.

However, based on the mass-X-ray luminosity  relation, the upper limit on the 
X-ray luminosity of \rhalo\ translates to an  
upper limit for the gravitational mass of its host cluster/group of
about $ 10^{14}$ $M_{\odot}$ (e.g., \citet{RB02,Stanek06}). 
This suggests that \rhalo\ is possibly located at the centre 
of a poor-cluster or a rich-group
environment, which is also  corroborated by the (photographically estimated) 
low galaxy density reported by \citet{Lopes04} for the galaxy 
cluster NSCS~J011904+112133, which hosts \rhalo. This is a rather 
surprising result which makes \rhalo\ special because 
in general, all other known mini radio-haloes are found within X-ray bright
cooling cores (Table~3) with X-ray luminosity \gtsim$10^{44}$ erg s$^{-1}$, and
some of the most powerful systems may have luminosity $\sim$10$^{45}$ erg s$^{-1}$.

%an important fact and makes this object special.
%We should perhaps use in the abstract more superlatives and say that  
%this seems to be the lowest mass system with a mini-radio halo ever  
%observed but yet one of the most radio-luminous ones.

\subsection{Spectral index plot: The `break' frequency and   radiative ageing
analysis}
As  these 100~kpc scale  giant radio-emitting magnetized bubbles  in \rhalo\  suggest their origin in intermittent AGN activity, 
we can estimate  the timescale of such outbursts from the shape of
total radio spectrum and the spectral index maps. 
Spectral index represents a powerful tool to  study the 
properties of the relativistic electrons and  the magnetic 
field in which they radiate, and to investigate the  connection 
between the electron energy and the ICM. 
From our multi-frequency radio imaging  and published data, we have derived the 
integrated spectrum (Fig.~\ref{rad_spectr}) and
spatially resolved spectral index maps (Fig.~\ref{maps_spectindx}).
Low frequency ( $\sim$100 - 400 MHz) spectra are  important to 
determine the slope of electron 
energy distribution and preserve a 
record of past AGN activity,  while the high
frequency spectra ($\sim$1 - 10 GHz) give information on the
diffusion and ageing of relativistic particles, and any particle
(re)acceleration  mechanism that may be in operation. 

The integrated radio spectrum 
between 74~MHz and 4.9~GHz frequencies  shows a
strong downward curvature (Fig.~\ref{rad_spectr}).
Piecewise linear least-squares  fit shows an
onset of  spectral steepening beyond
the critical (also known as `break') frequency $\nu_{c} \sim$400 MHz. 
The low-frequency spectral
index is $\alpha_{l}=-0.55(\pm0.05)$, while the
high-frequency spectral index attains a slope $\alpha_{h}=-1.35(\pm0.06)$. 
The steep spectral index of \rhalo\  is consistent with   
the mini radio-halo morphology which usually have steep radio spectrum 
($\alpha<$ -1, \citet{Ferrari08}). 
If the relativistic electrons are all injected in a single energetic
outburst with a power-law energy distribution, $N(E) dE = N_{0} E^{-p} dE$
(here $N_{0}$ is amplitude and $N(E) dE$ the differential number density
within energy $E$ to $E+dE$, and $p$ the power law index), subject to
strong radiation losses in synchrotron and inverse Compton processes, the radio 
spectrum would cut-off rapidly to zero for frequencies above
a critical frequency $\nu > \nu_{c}$, which shifts to lower values with time. In contrast, 
if episodic injection of fresh particles, or re-acceleration of existing particles takes place in the source, 
only a spectral steepening by $\sim$0.5 beyond the critical frequency occurs.  
In \rhalo\  beyond the critical 
frequency $\nu_{c}$ the radio spectrum steepens due to radiative
losses, but no rapid cut-off is seen (Fig.~\ref{rad_spectr}), 
which suggests an episodic or multiple particle 
injection process rather than a single energetic outburst event.
%The injection of relativistic
%particles in the medium can take place via an episodic on-off mechanism of 
%the radio-jets from the central AGN, or via a shock/turbulent 
%re-acceleration mechanism. The requirement is that $t_{sp}$ should
%be longer than the time interval $t_{int}$ between two successive radio
%outburts or than the acceleration time scale $t_{acc}$ for emitting electrons, 
%i.e. $t_{sp} > t_{int}$ or $t_{sp} > t_{acc}$.

The spectral age $t_{sp}$ derived from the synchrotron radio spectrum
is given by \citep{Slee01}:

\begin{equation}
t_{sp} = 1.59 \times 10^9 \frac {B^{0.5}}{(B^2+B_{CMB}^2)(\nu_{c}
(1+z))^{0.5}} \,\, ~yr,
\label{spect_age}
\end{equation}
where the magnetic field B is in $\mu$G,
the critical frequency $\nu_{c}$ in GHz and 
$B_{CMB} = 3.2 \times (1+z)^2 \mu$G is the 
equivalent `magnetic field'  of the cosmic microwave background (CMB)
radiation at redshift z. This formula  assumes isotropic pitch-angle
distribution, no expansion losses and a 
uniform magnetic field which remained unchanged over the radiative age. 
For $\nu_{c} = 400$ MHz and  magnetic field range B$= 1 - 10$ $\mu$G, the
electron spectral age is $t_{sp} = (1.33 \,-\, 0.64) \times 10^{8}$ 
y, which can be taken as the elapsed time since
 last injection of relativistic particles in the source. 
The largest uncertainty in estimation of spectral age is 
imposed by magnetic field B.  Presently we do not
 have X-ray observation of this source, but from the observed mini 
 radio-halo morphology, it is likely that \rhalo, in common
with other known mini radio-haloes, would  be located at the centre of a 
cluster cooling-core, which in general 
have stronger ICM magnetic fields of order B$= 10 - 30$ $\mu$G compared 
to non cooling-core clusters \citep{Govoni04,SS90}. Therefore 
$t_{sp}\sim$few$\times10^{7}$ y is a more likely (globally
averaged) radiative time-scale for \rhalo. 

The injection of relativistic
particles in the medium can take place via an episodic on-off mechanism of
the radio-jets driven by the central AGN (even though no
radio core or jets are visible down to $\sim$1 mJy/beam level), or via a 
Fermi-type shock/turbulent re-acceleration mechanism. The 
requirement is that cooling time
$t_{cool} \approx t_{sp}$ should
be longer than either the time interval $t_{int}$ between 
two successive nuclear outburst, or the acceleration 
time-scale $t_{acc}$ for  electrons emitting
at frequencies $\nu$ \gtsim $\nu_{c}$, i.e., $t_{sp} > t_{int}$ 
or $t_{sp} > t_{acc}$. 
It should be added that a lack of spectral cut-off upto 
the highest observed frequency $\nu$ = 4.86 GHz, 
where electron cooling time-scale  $t_{cool} \approx 1.8 \times 10^{7}$ y 
(B = 10 $\mu$G), puts a severe constraint on the acceleration 
time-scale, which should 
be shorter than this, irrespective of the acceleration mechanism.

%\label{scalings}
\begin{table*}
%\centering
\begin{minipage}{84mm}
\caption{{Comparison of \rhalo\  with other known mini radio haloes}$^{\dag}$.
Column 1: Cluster name. Column 2: Cluster redshift. Column 3: Log X-ray
luminosity of the cluster in energy range 0.1-2.4 keV. Column 4: Log 1.4 GHz 
radio luminosity.
Column 5: Log radio halo radius in kpc.}
%\begin{center}
\label{Tab.minihalo}
\begin{tabular}{lrrrr}
\hline
\hline\noalign{\smallskip}
Name   & Redshift &  Log $L_{\rm X}$ & Log $L_{\rm 1.4\,GHz}$ & Log $R_{\rm H}$\\
       &   &  (erg s$^{-1}$)& (W Hz$^{-1}$) & (kpc)\\
\noalign{\smallskip}
\hline\noalign{\smallskip}
       Perseus& 0.018 & $44.82$ & $24.27$ & 2.12 \\
       A2390& 0.228 & $45.13$ & $24.99$ & 2.26 \\A2626& 0.060 & $44.03$ & $23.36$ & 1.85 \\RX J1347.5$-$1145  & 0.451 & $45.65$ & $25.28$ & 2.41 \\
Z7160& 0.258 & $44.93$ & $24.34$ & 2.24 \\
RBS 797& 0.350  & $45.31$ & $24.63$  & 2.01 \\
M87&0.004   &$43.90$  &$24.72$   &1.60  \\
\rhalo& 0.131  &$<43.15$$^{\star}$  & $24.66$  & 2.08 \\
\noalign{\smallskip}
\hline\noalign{\smallskip}
%\footnotetext{$\dag$ Data from \citet{Cassano08} and \citet{Birzan08}.}
\end{tabular}\\
%\end{center}
$\dag$ Data from \citet{Cassano08} and \citet{Birzan08}\\
$\star$ Upper limit estimated by us from {\it ROSAT All Sky Survey} (RASS) data 
in 0.5 -- 2 keV band
%(a) Sijbring 1993,
%(b) Bacchi et al. 2003,
%(c) Gitti et al. 2004,
%(d) Gitti et al. 2007,
%(e) Venturi et al. 2008,
%(f) Gitti, Feretti \& Schindler 2006, Gitti et al., in prep.
\end{minipage}
\label{Tab2}
\end{table*}

\subsection{Spectral index maps: particle re-acceleration and a buoyantly 
rising plasma bubble?}
\begin{figure*}
\centerline{\psfig{figure=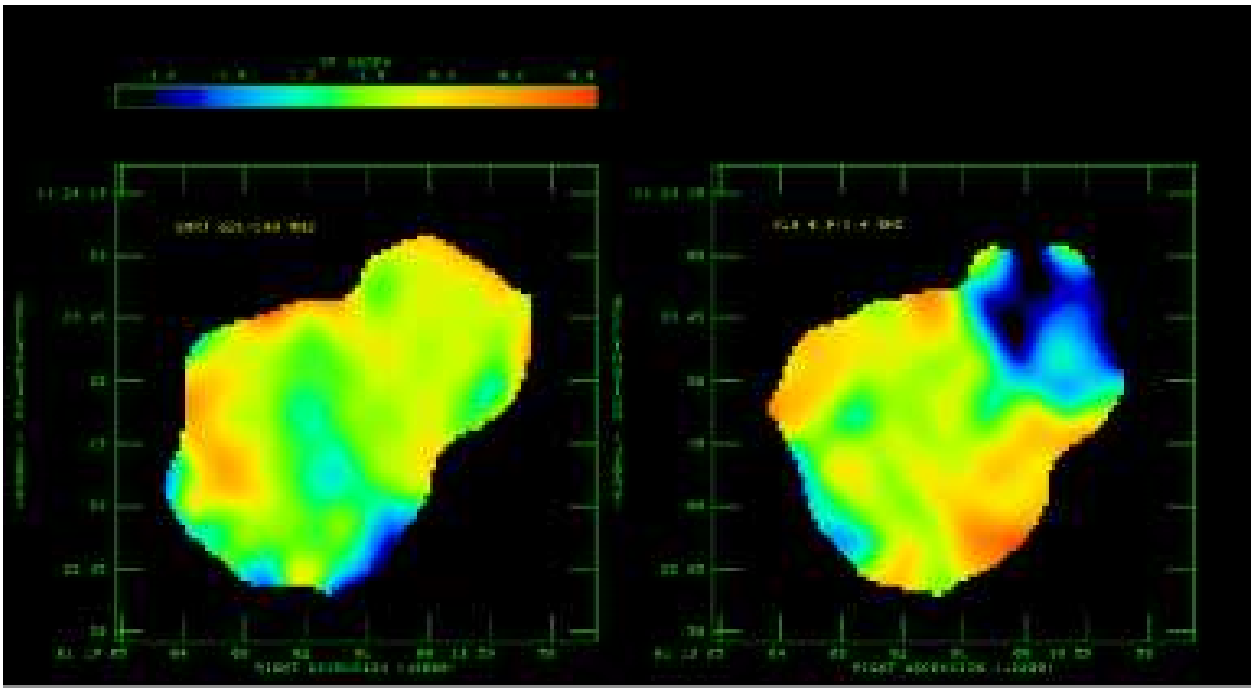,angle=-90,width=16.6truecm}}
\caption{Low and high frequency spectral index maps
of \rhalo\ obtained by combining the GMRT radio
maps at 240 and 621 MHz ({\bf left panel}) and
VLA maps  at 1425 and 4860 MHz ({\bf right
panel}). Both pairs of maps have the  matched resolution of 12\arcsec\, (FWHM).
To prevent spurious structures  from appearing, only pixels above $\sim$3.5 times
the rms noise level were included by giving  cut-offs
at values 4 mJy/beam, 0.5 mJy/beam, 0.27 mJy/beam and 0.15 mJy/beam
at 240, 621, 1425 and 4860 MHz  respectively.
The colour-bar shows  the spectral index values.
A significant spectral steepening (mean $\alpha = -1.6$)
for the north-western bubble at high-frequency can be clearly seen.
}
\label{maps_spectindx}
\end{figure*}

Here we present the spatially
resolved spectral index maps which reveal 
interesting details of the radiative
energy loss, particle re-acceleration processes and  bubble dynamics 
in this radio source. The spectral slope of radiation
emitted from different parts of the source is indicative of radiative ageing (or
lack of it) of emitting particles.
The low and high frequency spectral index maps are presented in 
Fig.~\ref{maps_spectindx}. These maps 
show that between 240 and 621 MHz the spectral index distribution
is fairly uniform with no strong steepening or gradients across the 
source ($\alpha_{mean} = -1.02\pm0.17$ for south-western bubble and
$\alpha_{mean} = -0.95\pm0.10$ for the north-eastern bubble, 
which are marked in Fig.~\ref{gmrt_2}),
implying  a lack of strong radiative losses 
between these frequencies (Fig.~\ref{maps_spectindx}, left panel).
However, the high frequency
spectral index map  between 1.4 and 4.8 GHz
presents a strikingly different picture (Fig.~\ref{maps_spectindx} right panel).
On this map, while the south-eastern bubble still has the same average
spectral index value around -1
($\alpha_{mean} = -1.06 \pm{0.15}$) - implying a straight synchrotron spectrum
between low and high radio frequencies - the north-west bubble, in contrast has
developed an extremely steep spectrum ($\alpha_{mean} = -1.6 \pm{0.20}$), most
likely due to  strong radiative energy losses in this part of the source.

Such a situation might  be explained if we assume that
the north-western bubble showing the very steep spectrum
is buoyantly rising away and then detaching itself from a centrally located
mechanism of  energy injection, while this source or mechanism is
still active and possibly injecting  relativistic particles
into the relatively flatter spectrum south-eastern bubble. One possibility 
is AGN related activity connected to a super-massive nuclear black hole, 
which is suggested by 
the  large bolometric radio luminosity of 
\rhalo\ ($L_{\rm radio}$ = $3.64 \times 10^{34}$ W, see 
Table~2), though we fail to detect a nuclear core or 
active jets (Fig.~\ref{gmrt_1}). It is plausible  that 
these bubbles are tracers of a previous cycle of AGN activity. Thus, 
a possible scenario is 
that these bubbles   were  inflated  by the 
pressure of radio-jets in a previous very
energetic episode of AGN  activity which has either stopped now or
has become too feeble to be detectable. During the later stages of evolution the 
dynamics of the remnants of the radio-jets  
will be dominated by buoyancy forces 
and drag (e.g. \citet{GN73}, \citet{Churazov01}), as they interact with the
cluster atmosphere in the inner core region. 
Our detection of a possibly buoyantly rising plasma bubble is  
also significant from theoretical
point of view, as one of the most appealing and 
widely applied models in literature 
for heating of ICM involves buoyant gas bubbles, inflated by AGN source, 
that subsequently rise through the ICM and 
heat it up (e.g.,   \citet{Churazov02,Bohringer02,BK02,Bruggen02,Cattanoe07,Ruszkowski08}).  
We further discuss the physics of these radio bubbles in the 
sections below.

\subsection{Remarkable morphological similarity with the twin radio-bubbles
of M87 (Virgo A)}
Our radio observations presented above clearly show that
\rhalo\  is a striking example of a radio-halo  source consisting of twin
plasma bubbles, possibly inflated in  a previous episode of AGN outburst
from  a super-massive black hole. We wish to point out that, in this respect, 
it  closely resembles
the radio emitting bubbles observed around M87,
the dominant central radio galaxy of the Virgo cluster \citep{OEK},
3C84, the central cD in Perseus cluster \citep{pedlar90} and 3C218
the central radio source in Hydra cluster \citep{McNamara_et_al2000,Wise07}.
However, as we have pointed out above, a notable major difference is
that while these (and other cluster centre radio mini-haloes, 
see Table~3) exhibit the
central galaxy's active nucleus (central core) and visible
jet structure,  \rhalo\  in contrast is probably a quiescent  source,  lacking
an active AGN and  ongoing radio-jet activity. Therefore,  
\rhalo\  is quite unusual and presents a
unique opportunity to understand the dynamical 
evolution and impact of  radio jet inflated
plasma bubbles on the ICM, when the central AGN  activity approaches cessation (probably when the fuel supply to the SMBH is exhausted), or when the non thermal 
bubbles start rising buoyantly and detaching themselves from the flow.

\begin{figure}
\psfig{file=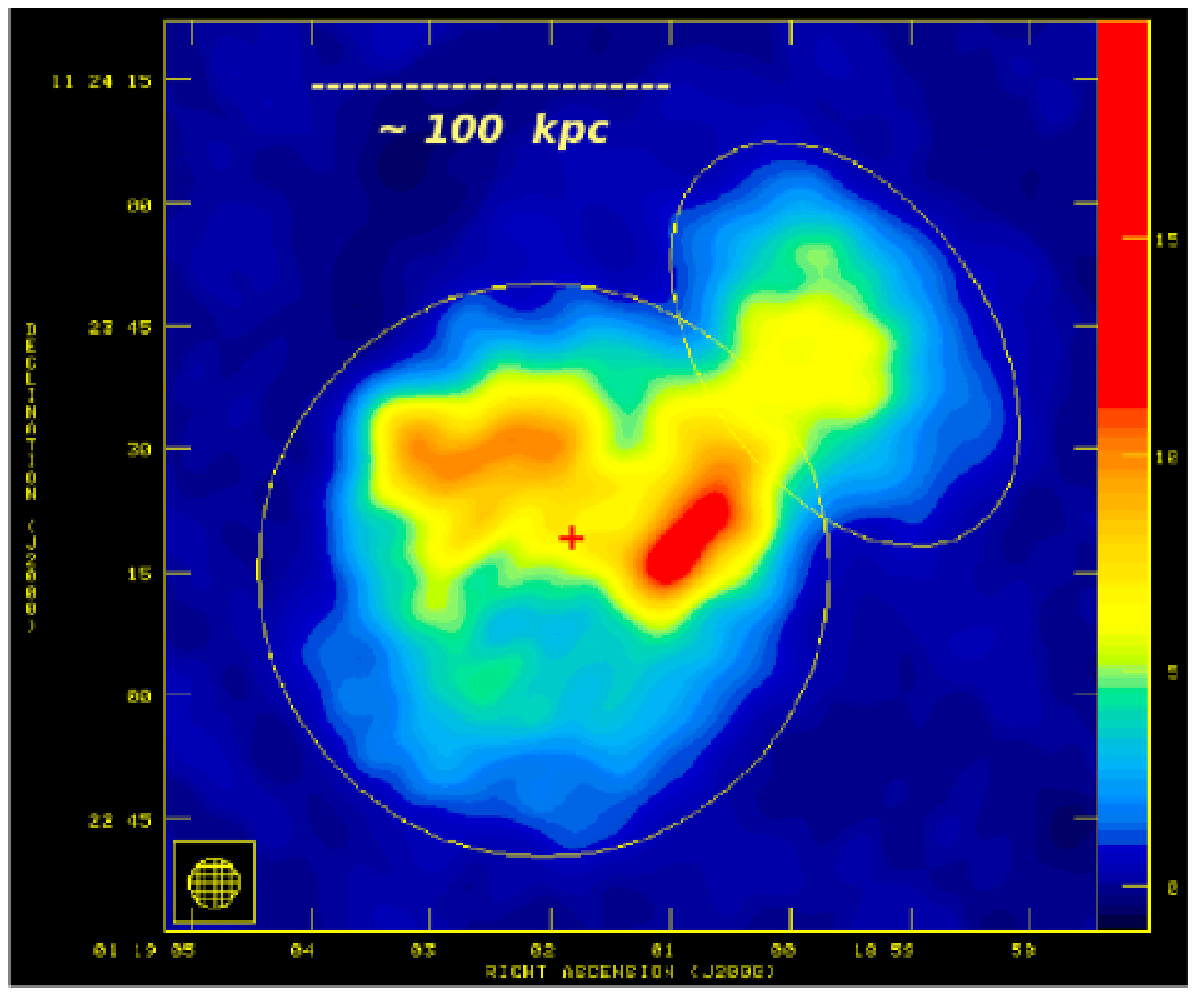,width=8.3truecm}
\psfig{file=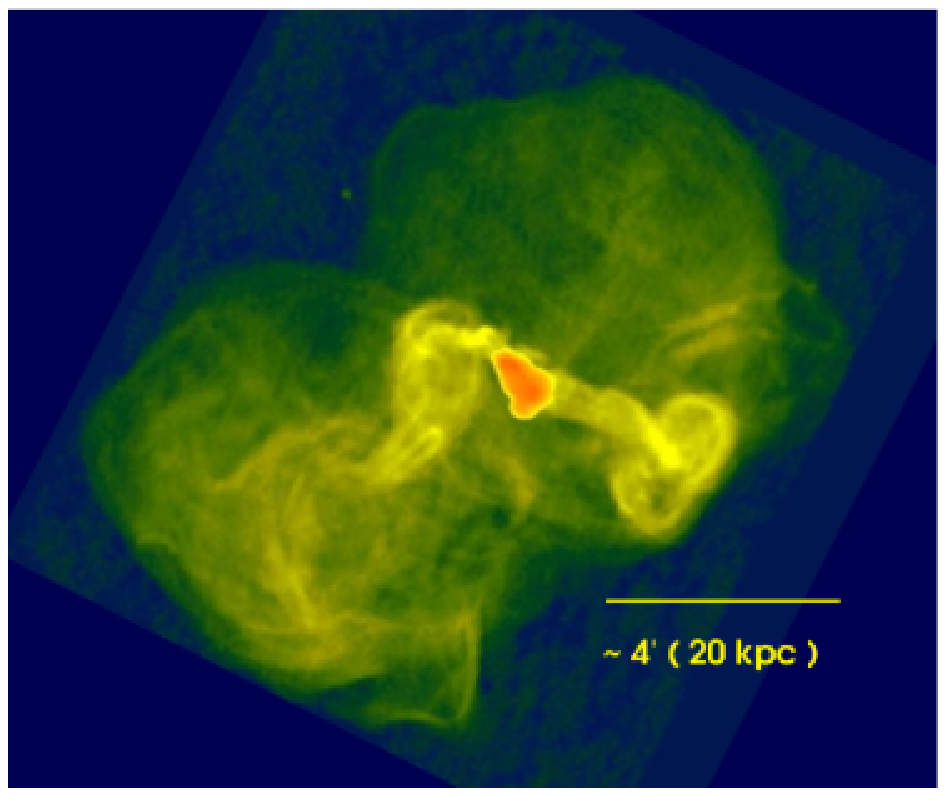,width=8.3truecm}
\caption{{\bf Upper figure:} GMRT 621 MHz false-colour image  of \rhalo
(beam: 6\arcsec\ FWHM circular,  plotted inside the box) showing the twin bubble-like diffuse
radio structure. The color-bar shows surface brightness in mJy/beam units and
position of brightest central galaxy (cD) is shown by a {\bf `+'} symbol.
The dotted lines delineate the two radio bubbles having
projected dimensions $\sim$27.5\arcsec $\times$17.5\arcsec
(semi-major and minor axes
of ellipse) for the smaller elliptical north-west
bubble and $\sim$35\arcsec (circular radius)
for the spherical larger bubble south-east of the cluster centre.
{\bf Lower figure:} The VLA 90~cm  radio image
of M87, the dominant central radio galaxy in the Virgo Cluster \citep{OEK}. 
The original map  has been reflected and 
rotated for a better comparison with the \rhalo\  radio map (reproduced by permission of the 
AAS).
}
\label{gmrt_2}
\end{figure}

\begin{figure}
\psfig{file=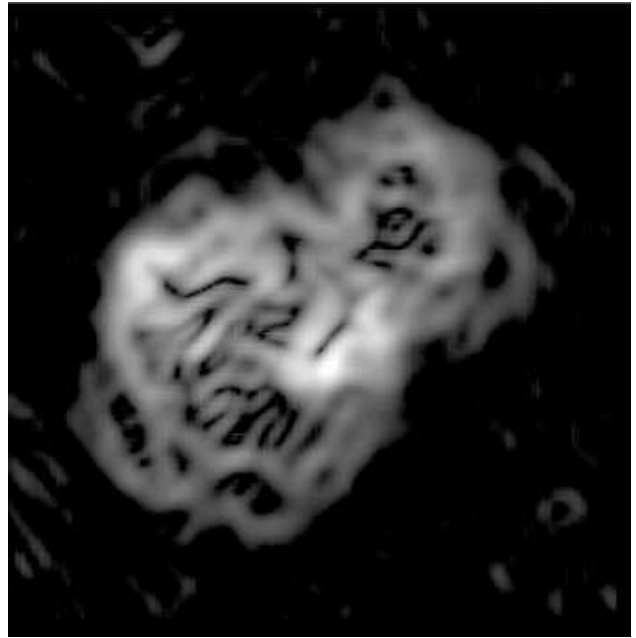,width=8.3truecm}
\caption{An image processed view of  GMRT 621 MHz image of \rhalo\  
shown in Fig.~\ref{gmrt_2}. We have used the   `Sobel' gradient filter operation
as implemented in AIPS task NINER to enhance sharp intensity
gradients and suppress diffuse emission. Highly intricate filamentary structures of
radio emitting plasma interior to the radio-halo can be seen.
}
\label{gmrt_filter}
\end{figure}

For a closer comparison, in Fig.~\ref{gmrt_2} we show the 621 MHz
radio image of \rhalo\  and a
deep VLA 90~cm  image of  M87 published by
\citet{OEK} (the original image has been reflected  and rotated for
better comparison with \rhalo). The morphological similarity between
the two radio sources is striking. \citet{OEK} found that in M87,
beyond the well known inner-jet and lobe region
(which appears as the 2~kpc scale red-orange patch near the centre of M87,
Fig.~\ref{gmrt_2}), there is diffuse outer structure
extending upto $\sim$40 kpc from the nucleus, which is
known to host a central supermassive black hole with a mass of 
$3.2\times 10^9~M_{\odot}$ \citep{Harms94}.
Two large scale collimated flows emerge from
the inner-jet region, one directed  north-eastward and the
other oppositely directed to south-westward (orientations refer to 
the rotated image). Particularly intriguing is a well-defined pair of edge-brightened, `ear-shaped' torus-like
structure at the end of south-west flow, which is reminiscent of a subsonic vortex ring. The north-eastern flow develops a gradual but well-defined S-shaped
southward twist, starting only a few kiloparsecs beyond the inner lobes.
Finally, both flows are  embedded in a giant diffuse radio structure
that might be described as two overlapping `bubbles', each extending
about 40 kpc from the nucleus. After reaching the halo, the flows
gradually disperse, particularly the north-eastward flow, and appear to be
filling the entire halo with radio-loud, filamented plasma.

%Hydrodynamic simulation results of \citep{Churazov01} suggest that
%these bubbles in M87 are buoyant bubbles of cosmic rays, inflated by an
%earlier nuclear active phase of the galaxy, which  rise through the
%cooling gas at roughly half the sound speed.

We observe similar flow pattern in the central region of \rhalo, despite 
the limited spatial resolution available. As shown in Fig.~\ref{gmrt_2}, an edge brightened torus-like `mushroom' structure
is observed about 40~kpc west
of centre of the larger south-eastern  bubble, which might well 
be  analogous to the peculiar `ear-shaped' vortex
structure of M87. Here the flow pattern sharply turns northward and
appear to be flowing into the smaller radio bubble to the north-west.
We observe an S-shaped flow pattern to
north-east of centre, which
further bends to south, and clearly resembles the  filamentary
structure found  in the southern bubble of M87. In order to better visualize
a  possible filamentary structure of synchrotron plasma bubbles, we have digitally
image processed the  621 MHz map of \rhalo\  with the AIPS task NINER 
for a gradient filter operation, which enhances sharp intensity
gradients and suppresses diffuse emission. The resulting image 
reveals (Fig.~\ref{gmrt_filter}), akin to M87, 
a highly complex filamentary structure of 
plasma embedded in the radio-halo. However , with the 
available angular resolution (6\arcsec\, FWMH beam size) 
far less details are visible in \rhalo\  as compared to M87, mainly
because it is much farther away from us than M87.
Of course, as mentioned above, in our radio observations of \rhalo\  
we find no evidence for a strong AGN core, inner-jets and
lobe structures, which clearly are present in M87.
One feature which makes the case of \rhalo
especially interesting is that the inner
plasma outflow structures in both these radio sources
are surrounded by a pair of diffuse radio emitting `bubble' shaped giant lobes.
We note that large misalignment seen in these two cases, between the inner 
outflow structure and the surrounding mini-halo may in fact be fairly common,
exhibited by several other mini-haloes. Examples include A2626 and the
prototypical mini-halo in Perseus-A \citep{Gitti04,pedlar90}. All this hints towards a common mechanism, namely the radio-jets from AGN, strongly interacting with the confining environment of a cooling core, forming an amorphous mini-halo structure.

Moreover, it is also interesting to note  that
although \rhalo\  closely resembles M87 in its radio
morphology, at $\sim$240 kpc its total linear size is about
three times larger than M87, which spans a scale of
just $\sim$80 kpc (\citet{OEK} and Table~2).
The bolometric radio luminosity of both
is about the same;  $L_{\rm radio}$ = $3.64 \times 10^{34}$ W
for \rhalo\  and $L_{\rm radio}$ = $3.6 \times 10^{34}$ W for M87
\citep{Birzan08}. This in turn implies that the average 
volume emissivity of \rhalo\  is almost 30 times smaller than that of M87. 

\section{Discussion}

The similarity of  large scale radio
plasma outflow structures observed in \rhalo\
and M87 suggest that both sources might have originated in,
and their evolution governed by, similar physical process and conditions 
prevailing in the central regions of their host clusters.  The origin
of  extremely complex diffuse radio structures observed in the 
outer radio bubbles of M87 and other similar radio mini-halo objects is not 
yet fully understood. From Hydrodynamic simulation 
\citet{Churazov01} suggested that the twin bubbles in M87 are  
buoyant bubbles of cosmic rays and magnetic fields, 
inflated by jets launched during an earlier nuclear 
active phase of the central galaxy,
which  rise through the cooling gas at roughly half its sound speed.
On the contrary, \citet{OEK} favor a scenario in which the 
radio halo of M87 is not simply a relic of previous episode of 
AGN activity but presently it is `alive', such
that it is being supplied with relativistic  particles  coming 
from the central AGN and the innermost radio jets. 
As revealed in our radio observations,  non-detection of a
strong AGN core, inner-jets and lobe/hotspot structure in \rhalo, 
coupled with the detection of a $\sim$100 kpc scale radio bubble 
of very steep spectrum north-west of the cluster centre supports the  `buoyantly-rising bubble' model of \citet{Churazov01}. The 
flattened `mushroom' shape of this plasma bubble indeed resembles a 
rapidly rising vortex-ring (Fig.~\ref{gmrt_2}), into which an 
initially spherical bubble will  naturally transform due to
viscosity and drag forces \citep{Churazov01,Bruggen02}. The present quiescent/low 
state of the central engine suggests an on-off activity cycle of the 
super massive black hole. That such activity is intermittent and 
the central AGN is periodically rejuvenated is suggested by
the high incidence of bubbles found in cluster cores that require heating 
to  counter balance the cooling flow \citep{Dunn_Fabian_2006}, 
and by the train of ghost-bubbles and pressure-ripples observed in Perseus and Hydra 
clusters \citep{Wise07,Sanders_Fabian07}. Also, the distant 
cooling core cluster  RBS797 shows an inner radio-jet pair 
oriented perpendicular to that of the extended 
radio-halo structure and X-ray cavities, suggesting re-starting activity of AGN \citep{GFS06}.

%According to our present understanding of the origin and
%evolution of mini-halos, their diffuse, steep-spectrum halo or bubble
%like morphology results from strong interaction of  radio jets with the
%surrounding dense and cool gaseous medium of the cluster core. 
%At present  there are no available good X-ray data for this system
%and therefore we lack information about the physical state of ICM  in the
%core of cluster hosting \rhalo. Such data is required for
%quantifying accurately the energy input rate by the AGN into the 
%plasma bubbles of mini-halo as well as their dynamics and interaction
%with the ICM. This cluster is located in an underexposed region of {\it ROSAT
%All Sky Survey} (RASS), which only allow us to determine an upper limit of
%1.4$\times$ 10$^{43}$ erg s$^{-1}$ for its luminosity in 0.5 -- 2 keV band. 
%In general, all the known mini radio-halos are found within X-ray bright
%cooling cores (Table~3) with X-ray luminosity $>$10$^{43}$ erg s$^{-1}$, and
% some most powerful systems may have luminosity $\sim$10$^{45}$ erg s$^{-1}$. 

An important question raised by our observations is: what 
energizes the relativistic particles in the 
central region of \rhalo\,  which apparently counter balances the 
strong radiation losses and  prevents severe steepening
of radio spectrum? The properties and physical processes governing the origin and
evolution of radio mini-halo sources are  poorly understood due to their
rather small numbers known till now (less than 10). 
Based on a sample of 5-6 mini haloes, \citet{Gitti04} have reported
a positive correlation between the radio power of  mini-halo and
the cooling flow radiative power. From this, and because of their estimate of the
lifetimes of the radiating electrons falling considerably short of the
diffusion time-scale, they have argued that in mini-haloes the 
cooling flow (through the
compressional work done on the ICM and amplification of frozen-in magnetic field)
energizes the particle acceleration process through magneto-plasma
turbulence acting on relic electron population probably left behind by
the past episodes of AGN activity. This suggests a direct connection between
cooling flows and radio mini-haloes. With a detailed spectral fitting of  the 
radio spectrum and emission profile 
with the parameters derived from the turbulent re-acceleration model,  it will be possible to test the viability of
this mechanism for the present source. 

An alternative proposal is that the radiating electrons in the mini-haloes are of 
secondary hadronic origin (via the pion decay route), created in proton-proton ($P_{CR}$-$P_{th}$)  
collisions \citep{Pfrommer_Ensslin04} of high energy cosmic ray (CR) protons ($P_{CR}$) with target thermal protons ($P_{th}$). The CR protons may originate in several ways: 
in large-scale structure formation shock waves 
\citep{Colafrancesco_Blasi_1998,Loeb_Waxman_2000,Ryu_et_al_2003,Bagchi_et_al_06}, 
or through supernovae and galactic winds 
from central  galaxies \citep{Volk_et_al_1996}, or  via powerful AGNs injecting  CRs into the ICM 
\citep{Ensslin_et_al_1997,Aharonian_2002,Pfrommer_Ensslin04}. A clear signature of 
this process is emission of $\gamma$-ray photons of broad energy range 
$E_{\gamma}$ $\sim$ 100 MeV - 10 TeV, accompanying  decay of neutral pions ($\pi^{0}$) 
produced in inelastic proton-proton collisions. Therefore, 
the hadronic model should be testable  in near future with sensitive  $\gamma$-ray observations. 
Recently, \citet{Hinton07} applied the hadronic CR injection model to  three galaxy clusters (Hydra-A, MS 0735.6+7421 and Hercules-A) harboring the most energetic AGN 
outbursts known (with mechanical energy input of AGN at least $10^{61}$ erg). They concluded  that the $\gamma$-ray observability of these objects is at the threshold of detectability with current (FERMI, HESS), and near future  $\gamma$-ray  instruments, if the observed 
radio bubbles are dominated by the pressure of relativistic particles. Therefore, such observations 
are obviously of great importance for deciphering the connection between the 
energetic CR and thermal energy content of clusters harboring the radio-haloes.

We have estimated the volume of the two radio bubbles from their projected 
dimensions on the GMRT 621~MHz image (Fig.~\ref{gmrt_2}; Table~2). The 
volumes are: $V_{1} = 4/3 (\pi r_{1}^{3}) = 6.78 \times 10^{70}$ cm$^3$
for the spherical south-west bubble of radius $r_{1} = 82$ kpc, 
and $V_{2} = 4/3 (\pi a^{2} b) = 2.08 
\times 10^{70}$ cm$^3$ for the ellipsoidal 
north-west bubble (we model it as an oblate
spheroid with semi-major axis $a = 64.4$ kpc and semi-minor axis $b = 41$ kpc). 
The  volume $V = V_{1} + V_{2} = 8.86 \times 10^{70}$ cm$^3$ is
the total volume occupied by relativistic particles and magnetic fields, 
which we assume to be mainly supplied by the radio jets and not 
produced in local shocks. The linear dimension of  the present radio bubbles
is $l_{bubble} \sim$ 100~kpc, which 
is much larger compared to the typical dimensions of X-ray cavities 
($l_{cavity} \sim$ 1 - 20~kpc; \citet{Birzan04}), and they  are comparable to the  
$\sim$100 -- 200 kpc size super cavities found in MS0735.6+7421 and 
Hydra clusters (\citet{McNamara_et_al2000,McNamara05,Wise07}). 

The large radio luminosity and linear size of the radio bubbles in \rhalo\
suggest that radio jets  from AGN must have done large amount of
mechanical work ($p dV$) in inflating them against the external  pressure of  
thermal ICM gas, as well as injected enormous non-thermal  plasma energy 
(magnetic field and relativistic particles), which has remained stored 
inside the bubbles, as they rose buoyantly under pressure equilibrium with the external medium. 
Future X-ray observations of \rhalo\  would reveal if the relativistic 
fluid filling these bubbles has  pushed aside the thermal gas, 
excavating  giant cavities in the X-ray emitting ICM. 
We have detected a  buoyantly rising radio bubble  $\sim$ 100~kpc from the
centre of radio-halo (Fig.~\ref{maps_spectindx}). Such bubbles 
result from displacement of the thermal gas, 
which creates a low-density  bubble maintained in pressure balance with the 
surrounding medium and rising subsonically. These non-thermal plasma bubbles
associated with X-ray cavities found in the ICM of clusters 
are turning out to be excellent `calorimeters' for guaging the 
total mechanical power of 
radio-jets from AGNs. Such measurements are extremely  useful as they not 
only preserve the past record of enormous outbursts of the  central SMBH, 
but  are also independent of the radio properties of the jets 
(i.e., their present radiative output and magnetic field strength), 
 allowing a good estimate of energy injected into the surrounding 
thermal plasma for heating it (e.g., \citet{Churazov02,Birzan04,Birzan08}). 

A thermodynamical measure of the total energy required to slowly inflate a hot cavity/bubble in ICM is 
given by its enthalpy, $ H_{cav}$=$[\gamma/(\gamma -1)] pV$, where $\gamma$ is ratio of specific heats of the gas inside the cavity, p is the pressure of external medium and V is the volume of the cavity \citep{Churazov02,Birzan04}. 
For a radio jet inflated non-thermal bubble,
which has completely displaced the thermal medium, 
$ \gamma = 4/3$ and $ H_{cav} = 4 pV$. The external 
pressure is $ p = \rho_{ICM} k T/(\mu m_{H})$ (where $ \rho_{ICM}$ is 
the density of ICM, T is the gas temperature, 
$\mu$=0.6 the mean molecular weight, and $m_{H}$ is  mass of the hydrogen atom), 
which is presently not known for \rhalo. We can obtain an  
order of magnitude estimate for $ H_{cav}$ if we take  fiducial
values, $\rho_{ICM} = 1.67 \times 10^{-26}$ gm~cm$^{-3}$ (for a proton 
number density of $10^{-2}$ cm$^{-3}$) and $T = 2.32 \times 10^{7}$ K (2 keV), 
which are typical of the gaseous environment around  a mini radio-halo 
source like M87 or Perseus-A at the centre 
of a cooling core \citep{Ghizzradi04}. This gives a pressure
$ p = 5.3 \times 10^{-11}$ dynes~cm$^{-2}$ and an
enthalpy $ H_{cav} = 2.06 \times 10^{60} V_{70}$ 
erg for a cavity volume $ V_{70} = V/(70 \, cm^{3})$. 

Using the measured volumes of the two radio bubbles in \rhalo\  (see above), we 
estimate  an enthalpy $ H_{1} = 1.39 \times 10^{61}$ erg for the south - east bubble, 
and $H_{2} = 4.28 \times 10^{60}$ erg for
the north - west `buoyant' bubble, resulting in  
total enthalpy $H_{1+2} = 1.82 \times 10^{61}$ erg. We require X-ray
data to estimate the external ICM pressure around the bubbles; but conservatively, even if 
this is 1/10 of the value assumed above, 
we get a total enthalpy  $H_{1+2} \sim 10^{60}$  erg which is still 
quite large, but for an  order-of-magnitude  consistent with the 
large radio luminosity and huge volumes of these bubbles. 
The  energy injection rate (mechanical luminosity) of radio jets,  
that inflated the radio bubbles, is $L_{j} \approx H_{1+2}/t_{j}$, 
where $t_{j}$ is the time interval for which jets were active out of the total
activity cycle time $t_{d}$ of the AGN. Various lines of 
evidence show that for radio galaxies
typically, $t_{d} \approx$few$\times 10^{8}$ y, and 
assuming $t_{j} \sim$few$\times 10^{7}$ y 
with $H_{1+2} \sim 10^{60}$ erg would lead to 
$L_{j} \sim$few$\times 10^{44}$ erg s$^{-1}$. \citet{Birzan08}  
derived an empirical relation (with considerable scatter);
log$\, L_{cav} = (0.35\pm0.07)$\,log$\,L_{1.4 \, GHz} + (1.85\pm0.17)$, connecting 
%$L_{j} = 1.2 \times 10^{44} [L_{\rm 1.4 \, GHz}/10^{25} {\rm W \ Hz^{-1}}]^{0.4}$ erg $s^{-1}$, by comparing 
the estimated mechanical jet luminosities of radio AGNs ($L_{j} = L_{cav}$) that have 
associated X-ray cavities, and their 1.4 GHz monochromatic radio luminosities. In this
relation $L_{1.4 \, GHz}$ is in units of $10^{24}$ W~Hz$^{-1}$ and the cavity (jet) luminosity $L_{cav}$ 
is expressed in units of $10^{42}$ erg~s$^{-1}$. 
From this relation, and with  $L_{\rm 1.4 \, GHz}$ = $4.57 \times 10^{24}$ W~Hz$^{-1}$ for \rhalo\ 
(Table~2), we obtain $L_{j} = 1.2 \times 10^{44}$ erg $s^{-1}$, which is 
consistent with the earlier `calorimetric' estimate. Moreover, adopting  $t_{j} = 10^{7} - 10^{8}$ y, 
the same as the synchrotron age derived in Sect.~3.3, we estimate that the 
total energy content (enthalpy) of the bubble sytem is
$\sim$\,4$\,\times\,(10^{58} - 10^{59})$ erg, which is less than our earlier
estimate ($10^{60} - 10^{61}$) erg, but we believe it represents best the situation in \rhalo. 
However, the exact figure can only be obtained from the
future X-ray observations of \rhalo.

Indeed,  both observations and theory suggest that the 
energy input from AGN into the cluster atmosphere  at 
this rate is  sufficient to heat it up and 
balance the cooling loss \citep{Rafferty06,Dalla_Vecchia04}. This may even halt 
(or reduce drastically) the accretion flow of matter onto the SMBH in some systems
(e.g., \citet{Cattanoe07}) - thus cutting off the fuel supply to the
`central engine' and completing the negative feedback loop. We wonder,
if this AGN mediated feedback process could also be
 at work in \rhalo, giving rise to its relaxed, 
quiescent looking appearance? If confirmed with future X-ray observations,
such large enthalpy of the bubble system in \rhalo\  would make 
it harbor one of the most energetic radio outbursts known, comparable to
the energetics of the super cavities in  MS0735.6+7421 and  
Hydra clusters \citep{McNamara05,Wise07}. 

The ultra steep spectrum north - west  bubble is located 
at a projected distance of R$\sim$100 kpc from the central cD galaxy, which is 
inferred to be buoyantly rising in the ICM. We do not know  how far back
in time it was inflated and where  was it injected into the ICM. Assuming that it was initially inflated rapidly by  an AGN-fed radio jet 
near the cluster centre, and subsequently  it detached and rose 
subsonically at $\sim 1/2$ the speed of sound 
$C_{s} = 717.6\,$$[T/(\rm 2 \, keV)]^{1/2}$ km s$^{-1}$, 
it will take $t_{s/2} = 2 C_{s}^{-1} R = 2.7 \times 10^{8}$ y 
for it to arrive at its present location (neglecting projection effects). 
This time-scale is much longer than the electron cooling time 
scale  $t_{cool} \approx 2 \times 10^{7}$ y (taking B = 10 $\mu$G) 
at  4.86 GHz, thus consistent with its observed 
very steep radio spectrum (Fig.~\ref{maps_spectindx}). 
However the actual rise time 
should be somewhat shorter than 
the above estimate when we take into account the initial rapid inflation and 
acceleration phase (due to forward momentum of the radio jet), at the end of which 
the bubble will attain a constant terminal 
speed $v_{term}$ and rise in pressure balance with the external medium. 
Another relevant time-scale is the `buoyant rise' time $ t_{buoy}$, 
which is the travel time for a bubble moving  with the terminal speed $v_{term}$. 
The X-ray data show that usually $ t_{buoy} \sim  2 \times t_{s/2}$ 
\citep{Birzan04,Wise07}.

Several galaxy clusters also show X-ray surface brightness 
depressions that have no obvious association with bright radio emission --
the so called `ghost cavities', such as those seen in
Abell 2597 \citep{Mcnamara01},  and the matched pair of cavities 
in Hydra and Perseus clusters \citep{Fabian2000,Wise07}. These 
depressions are thought to have been created by  AGN outbursts that occurred in the more distant past, 
but whose radio emission has faded over time due to radiation and
expansion losses (eg. \citet{EGK01}). 
An interesting possibility, suggested by our discovery of 
an ultra steep spectrum detaching radio bubble in \rhalo\  
is that, that such buoyantly rising radio bubbles are  precursors of  the
ghost cavities observed in many X-ray observations. Such ghost bubbles provide
important clues for understanding how AGNs heat cluster atmospheres.
A crucial issue is the survival of  detached radio bubbles upto large distances 
from the central source. The 
stability and dynamics of radio bubbles,  once the jets 
have turned off (as  in the present radio-halo \rhalo)  is a topic of 
great importance for the interpretation 
of X-ray observations of clusters. 
{\it Chandra}  observations show that some of the ghost cavities are 
able to reach $\sim$20 - 100 kpc distances from the central AGN, 
while it is  found in several
numerical simulations that hot thermal bubbles, once released into the
cluster atmosphere at the end of a terminating radio-jet, initially  
rise supersonically but  are rapidly fragmented 
by Rayleigh-Taylor (RT) and Kelvin-Helmholtz (KH) 
instabilities. To prevent their destruction, 
the role of magnetic field in stabilization of the rising bubbles was explored by
\citet{DeYoung03}. Numerical simulations (e.g., \citet{Ruszkowski08}) find that 
in an upward-rising cosmic-ray filled bubble, 
the magnetic field `draping' effect at the upper contact surface of bubble indeed
has a strong stabilizing effect which prevents cross-field diffusion of particles.
A substantial fraction of cosmic-rays can thus be confined inside the bubbles 
on buoyant rise time-scales ($\sim 10^{7 - 8}$ y) even when the 
parallel diffusivity coefficient 
is very large. Magnetic field stabilization of the radio bubbles 
in \rhalo\  is an interesting possibility worth exploring, as 
these are still intact quite far away from the central cD: $\sim$100 kpc
for north -- west bubble and $\sim$40 kpc for south -- east bubble, 
requiring some stabilizing mechanism for their survival. Detailed polarization mapping of 
these features is required to check if 
the theoretical  `draped' structure of protecting magnetic field does exist. 

Lastly, \rhalo, with its  dual bubble-like mini radio-halo morphology, provides  an excellent  test-bed for
understanding the  dynamics   of radio-jet inflated  buoyant plasma bubbles 
in the hot cluster atmospheres, and the role of AGN feedback  in transporting and mixing of 
metals in the ICM. Low-entropy, metal enriched cold gas may be  uplifted from the cluster centre 
along with the energetic wakes of AGN-driven plasma bubbles upto $\sim 10 - 100$ kpc scales, which 
undergoes heating and mixing with ICM via hydrodynamical turbulence. 
Recently, from HST observations  \citet{Fabian08} discovered an extensive network of $H_{\alpha}$ emitting
cold, apparently magnetically supported,  filaments surrounding the mini radio-halo Perseus-A.
They inferred that these cold filaments are dredged up from the centre of the galaxy 
by the radio-emitting bubbles, 
buoyantly rising  in the hot intra-cluster gas, and act as markers of the AGN feedback process. 

We expect that future more detailed radio,  X-ray, $\gamma$-ray and optical  emission line
studies of the galaxy cluster hosting \rhalo\  would
provide crucial data, that would allow us to gain a much better understanding
of the complex thermal and non-thermal plasma processes
taking place in the gaseous environment around this puzzling
yet extremely interesting radio-halo source.
\section{Summary and Conclusions}
\begin{enumerate}
  \item We have presented detailed radio and optical observations of  \rhalo, which clearly have   revealed a luminous,  steep-spectrum, diffuse radio-halo  source of $\sim$~240 kpc dimension   located at the centre of  a  X-ray weak, 
low-mass cluster of galaxies at redshift z = 0.131.
  
  \item On larger scales ($\sim$100 kpc) the radio-halo  emission
  comprises  of two  diffuse  `lobe'-like structures  surrounding the  central cD and  its associated group of galaxies,  which we have interpreted as being a pair of radio emitting giant   plasma bubbles, filled with relativistic   particles (cosmic rays) and magnetic field.
  
  \item We do not detect  any  ongoing AGN activity, such as a compact core or
  active radio jets feeding the plasma bubbles. Thus, it is possible  that 
  these bubbles are tracers of a previous cycle of AGN activity and  were  inflated  by the   pressure of radio-jets in a highly energetic episode of AGN  outburst, which has  either ceased by now, or  has become too feeble to be detectable.
  
  \item The physical size, radio spectral index and morphology of \rhalo\
  are broadly consistent with the properties of a cluster radio mini-halo,
 which are rare objects,  have low surface brightness and possess
  a steep spectral index ($\alpha$\ltsim\,-1). They are
  mostly observed at the centres of cooling-core clusters, which indicates that their  origin and evolution are closely linked to the energy feedback from the central AGN  via radio jets into ICM and its cooling/heating processes.
  The radio luminosity  of \rhalo\  at 1.4 GHz,  $4.57 \times 10^{24}$ W Hz$^{-1}$,
  and the bolometric radio luminosity (over 10 MHz - 10 GHz range), 
  $3.64 \times 10^{34}$ W, would place it amongst the most
  luminous radio haloes known.
  
  \item  The integrated radio spectrum  shows a sudden  steepening 
  beyond the `break' frequency $\sim$400 MHz. 
  The low-frequency spectral index  is $- 0.55(\pm0.05)$ and  the
  high-frequency spectral index is $- 1.35(\pm0.06)$. The electron spectral age 
  was derived to be  $(1.33 \,-\, 0.64) \times 10^{8}$ y, for magnetic 
  field values 1 - 10 $\mu$G.

  \item The pair of huge ($\sim$100 kpc scale) plasma bubbles  show
  marked spectral index differences. Between 240 and 621 MHz (GMRT) both bubbles
  show a similar mean spectral index $\alpha_{mean} \approx -1$, indicating a
  lack of strong radiative losses. However, whereas between 1.4 and 4.8 GHz (VLA)  the  south-eastern 
  bubble still has the same mean spectral index  of   $\alpha_{mean} = -1.06 \pm{0.15}$, 
  the north-west bubble, in contrast, has
  developed  an extremely steep spectrum $\alpha_{mean} = -1.6 \pm{0.20}$,
  clearly due to  strong radiative energy losses.  We argued that ongoing in-situ particle re-acceleration, 
  probably via    merger  induced shocks and/or  turbulence is taking place in the south-eastern bubble, 
  and the north-western  bubble is detaching and buoyantly rising away from the centre in the hot cluster atmosphere.  
  Such uprising, ageing bubbles  are potential precursors of the giant X-ray dark 
  cavities and could inject  $\sim10^{60 - 62}$ erg mechanical energy into the ICM for 
  heating it, as observed in   X-ray observations of several clusters.

  \item We discovered a remarkable morphological similarity between \rhalo\ and
   M87, the well known dominant central radio galaxy in the Virgo Cluster. Similar internal 
   plasma outflow structures and network of magnetic filaments surrounded by  
   a pair of giant outer radio bubbles are observed.  This suggests that both these sources might have originated in,
   and their evolution governed by, similar physical processes and conditions
   prevailing in the central regions of their host clusters. Inspite of the notable similarities,
   at $\sim$240 kpc, the linear size of \rhalo\  is about
   three times larger than M87. The bolometric radio luminosity of both
   is about same:  $3.6 \times 10^{34}$ W.  
   This in turn implies that the average volume emissivity of \rhalo\ is 
   roughly a factor of 30  smaller than that of M87 halo.

   \item From the known empirical correlation between the estimated mechanical 
   jet luminosities of  radio AGNs  that have  associated X-ray cavities, and 
   their 1.4 GHz monochromatic radio luminosities,
   %Taking reasonable fiducial values of gas temperature and 
   %pressure within the putative cooling core surrounding  the radio mini halo \rhalo, 
   we estimated the total enthalpy (free energy) of  bubbles in \rhalo, $\sim 10^{59 - 60}$ erg and the
  mechanical luminosity of radio jets which inflated them, to be $\sim10^{44}$ erg s$^{-1}$.
  Energy input from  a central AGN into the cluster atmosphere  at
  this rate is  sufficient to heat it up, drive a massive outflow from the AGN, 
  which may counter balance a possible cooling loss, and may even halt 
  (or reduce drastically) the accretion flow 
  of matter onto the supemassive black hole, thus cutting off the fuel supply powering the central
  engine and  completing the  feedback loop. This AGN mediated 
  feedback process in \rhalo,  could explain  its relaxed,
  quiescent looking appearance. If confirmed with future X-ray observations,
  such large enthalpy of the bubble system in \rhalo\  
   and its possible AGN-feedback cycle would make
  it host one of the most energetic radio outbursts known, comparable to
  the energetics of the super cavities in  MS0735.6+7421 and Hydra clusters.

  \item \rhalo, with its rare, twin bubble-like morphology of mini radio halo and possessing 
  other striking properties as revealed in our observations,  provides  an excellent  opportunity to
  understand the energetics and the dynamical  evolution of
  radio-jet inflated  plasma bubbles in the hot cluster atmosphere.  It also help in  
  understanding what possible   role  AGN-feedback plays in heating of ICM, transporting and mixing of 
  metals in the ICM, the evolution  of central
  galaxies  and the growth of nuclear supermassive black holes.

\end{enumerate}
\section*{Acknowledgments}
We thank the operations team of the NCRA--TIFR GMRT observatory
and IUCAA Girawali observatory. We specially wish to  thank Vijay Mohan (IUCAA) 
for providing extensive help during optical observation and data analysis and
Hans B\"ohringer (MPE) for providing us with an upper limit on the X-ray 
luminosity of the source from the {\it ROSAT All Sky Survey}.
J. Jacob, N. Wadnerkar, J. Belapure and A. Kumbharkhane 
thank IUCAA  for facilities and local  support during several academic visits.
Norbert Werner was supported by the
National Aeronautics and Space Administration
(NASA) through Chandra Postdoctoral Fellowship Award Number PF8-90056 issued
by the Chandra X-ray observatory Center, which is operated by the
Smithsonian Astrophysical Observatory for and on behalf of the
National Aeronautics and Space Administration under contract NAS8-03060.

\bsp

\label{lastpage}

\end{document}